\documentclass[pre,nofootinbib,twocolumn, showpacs, superscriptaddress,10pt]{revtex4-1}

\usepackage{packages}
\usepackage{notations}

\begin{document}

\setlength{\abovedisplayskip}{5pt}
\setlength{\belowdisplayskip}{5pt}

\title{Probing the large deviations for the Beta random walk in random medium}

\author{Alexander K. \surname{Hartmann}}
\affiliation{Institut f{\"u}r Physik, Universit{\"a}t Oldenburg, 26111 Oldenburg, Germany}
\author{Alexandre Krajenbrink}
\affiliation{Quantinuum, Terrington House, 13–15 Hills Road, Cambridge CB2 1NL, United Kingdom}
\affiliation{Le Lab Quantique, 58 rue d'Hauteville, 75010, Paris, France}
\author{Pierre Le Doussal}
\affiliation{Laboratoire de Physique de l'\'Ecole Normale Sup\'erieure, PSL University, CNRS, Sorbonne Universit\'es, 24 rue Lhomond, 75231 Paris, France}

\date{\today}

\begin{abstract}
We consider a discrete-time random walk on a 
one-dimensional lattice 
%in presence of a time dependent random environment, 
with space and time-dependent random jump probabilities,
known as the Beta random walk. 
We are interested in the probability that, for a given realization of the jump probabilities (a sample),
a walker starting at the origin at time $t=0$ is at position beyond $\xi \sqrt{T/2}$ at time $T$.
This probability fluctuates from sample to sample and we study the large-deviation rate function which characterizes
the tails of its distribution at large time $T \gg 1$. It is argued that, up to a simple rescaling, this rate function
is identical to the one recently obtained exactly by two of the authors for the continuum version of the model. That
continuum model also appears in the macroscopic fluctuation theory of a class of lattice gases, e.g.
in the so-called KMP model of heat transfer. An extensive numerical simulation of the Beta random walk, based on an importance
sampling algorithm, is found in good
agreement with the detailed analytical predictions. A first-order transition in the tilted measure, predicted to
occur in the continuum model, is also observed in the numerics. 
\end{abstract}


%05.40.-a: Fluctuation phenomena, random processes, noise, and Brownian motion 
%02.10.Yn	Matrix theory
%02.50.-r	Probability theory, stochastic processes, and statistics 

\maketitle

\section{Introduction}

The macroscopic fluctuation theory (MFT) \cite{BertiniMFT2015} provides a coarse grained continuum description 
of the fluctuations of the density and current \cite{DerridaMFTReview2007} for a broad class of 
discrete stochastic systems in one dimension with a diffusive scaling at large time. One important example is
the symmetric exclusion process, where particles perform symmetric jumps onto neighboring
unoccupied sites on a lattice. Another example is the Kipnis-Marchioro-Presutti (KMP) model \cite{KMP}, 
a lattice model where each site has an energy and whose the dynamics is described by a random exchange of energy between neighbors.
Upon introduction of an asymmetry or a driving, such as in the asymmetric 
exclusion process \cite{DerridaReviewASEP}, the diffusive scaling breaks down above some scale, 
and the large scale behavior of the model is usually described by the Kardar-Parisi-Zhang (KPZ) universality class
\cite{TWASEP2009}. It was shown that there is a natural crossover from the MFT to the so-called weak noise theory (WNT)
of the KPZ equation as the asymmetry is increased \cite{KrajLedouCrossover}. 

The MFT and the WNT allow to reduce the calculation of the large deviations of density and current 
to solving a system of two coupled non-linear differential dynamical equations, with prescribed boundary conditions
both at initial and final time. Recently, starting with the WNT for the KPZ equation \cite{UsWNT2021,UsWNTFlat2021,TsaiDroplet2022}, 
exact solutions to these systems were obtained 
\cite{NaftaliDNLS,KrajLedouCrossover,grabsch2021closing,mallick2022exact,NaftaliDNLS2,GrabschBenichou2023Review}.
This was achieved by using the close connection of these systems to the non-linear Schrodinger
equation (NLS), or to the derivative NLS equation (DNLS),  
and extending the inverse scattering methods of \cite{ZS,AblowitzKaup1974,kaup1978exact} 
to mixed-time boundary conditions. Another largely equivalent method used exact closure schemes 
\cite{grabsch2021closing,GrabschBenichou2023Review}. This allows one to compute large deviations for observables such as the integrated current or of the position of a tracer.

Here we will focus on the case where the MFT takes the form of a linear stochastic
equation for a space-time coarse-grained density field $q_\eta(y,\tau)$
\be 
\label{FP} 
 \partial_\tau q_\eta(y,\tau) = \partial_y^2 q_\eta(y,\tau) - \partial_y ( \sqrt{2} \eta(y,\tau) q_\eta(y,\tau) )\, ,
\ee
where $\eta(y,\tau)$ is a standard space-time Gaussian white noise. It was proved in
\cite{bertini2005large} 
that at large time the large deviations for the discrete KMP model are identical to those of the continuum stochastic model \eqref{FP}.
At large time the dynamical action associated to model \eqref{FP} is controled by a saddle point, and the
corresponding saddle point equations define the MFT for this model. These MFT equations were 
studied in a number of works %and their solutions mostly by numerical methods, perturbation theory or asymptotic expansions 
\cite{BertiniPRL2005,DerridaGershenfeld,Lecomte,KrapivskyMeerson,Zarfaty,BodineauDerrida,BertiniMFT2015,Tailleur2007,Hurtado,Peletier,Shpielberg,grabsch2021closing,poncet2021generalized}. 
We noted in \cite{KrajLedouCrossover} an interesting
connection to a continuum model of diffusion in a time dependent random environment, previously
considered in \cite{TTPLD,BarraquandSticky,BarraquandBetaHalfSpace,WarrenSticky,WarrenEdgeCloud,das2023kpz}.
Indeed, Eq.~\eqref{FP} can also be seen as the Fokker-Planck equation for the probability distribution function (PDF) 
$q_\eta(y,\tau)$ of the
position $y(\tau)$ at time $\tau$ of a particle convected by the random field
$\eta(y,\tau)$, described by the Langevin equation
\be \label{Langevin} 
\frac{\rmd y(\tau)}{\rmd \tau}= \sqrt{2} \eta(y(\tau),\tau)+ \chi(\tau) \, ,
\ee
where $\chi$ is a standard white noise in time. The subscript in $q_\eta$ emphasizes
that it depends on the realization of the random field $\eta$, i.e. the sample. 
In \cite{KrajLedouCrossover} (see also 
\cite{NaftaliDNLS,NaftaliDNLS2}) we solved the MFT equations and derived the large-time large-deviation function 
associated to $q_\eta$ for the continuum model \eqref{FP},
with applications to diffusion of extremes in time-dependent continuum random media. This solution was
obtained by inverse scattering methods on a non-linear system interpolating between the DNLS and NLS
equations. 
We also obtained the same result by performing the large-time expansion of an exact Fredholm determinant formula 
obtained in \cite{BarraquandSticky} using the Bethe ansatz.

It is thus natural to investigate whether 
the MFT associated to \eqref{FP} can describe a discrete model of a random walk
in a random environmnent (RWRE). The natural example which we will consider here is the so-called Beta random
walk, introduced and studied in \cite{BarraquandCorwinBeta}. This model was later studied 
in relation to the KPZ equation in 
Refs.~\cite{TTPLD,CorwinGu,TTPLDBeta,BarraquandSticky,GBPLDModerate,WarrenEdgeCloud,das2023kpz}.
In the present paper we will first argue that the large-time large-deviations tails for
the Beta polymer are indeed identical, up to some simple rescaling
that we can predict, to those of the continuum model. Next,
we will perform an extensive numerical study of these large deviations
for the Beta random walk, using an importance
sampling algorithm, to test our analytical predictions.

The outline is as follows. In Section \ref{sec:model} we define the
model of the Beta random walk, introduce the observables of interest
and define the associated large-deviation rate functions. 
In Section \ref{sec:prediction} we sketch the argument which allows to
relate the large deviation of the discrete model to those of the
continuum model. In Section 
\ref{sec:methods} we explain the numerical method used here,
notably the importance sampling method which allows to explore the deep tails of the large-deviation
regime. In Section \ref{sec:compare} we give the main numerical results and discuss
how they compare to the analytical predictions.

\section{Model and observables} 
\label{sec:model} 

The model of the Beta random walk is defined as follows \cite{BarraquandCorwinBeta}. 
One defines first the "environment" or sample, by choosing for each $x \in \mathbb{Z}$
and $t \in \mathbb{N}$ a variable $w_{x,t} \in [0,1]$. The $w_{x,t}$ are
chosen as i.i.d random variables taken from the beta distribution with parameters $\alpha, \beta >0$ and density
\be 
{\cal P}(w) = \frac{\Gamma(\alpha+\beta)}{\Gamma(\alpha) \Gamma(\beta)}  w^{\alpha-1} (1-w)^{\beta -1}
\label{eq:beta:distr}
\ee 

\begin{figure}[h!]
    \centering
    \includegraphics[width=0.7\linewidth]{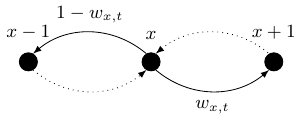}
    \caption{Lattice representation of the Beta random walk.}
    \label{fig:beta-random-walk-lattice}
\end{figure}

One now considers a particle at position $X(t) \in \mathbb{Z}$, 
which starts at the origin at time zero, $X(0)=0$, and
which performs a random walk defined by the following
transition probabilities (see Fig.~\ref{fig:beta-random-walk-lattice}) 
\be
\label{eq:move:probs}
\begin{split}
& \mathbb{P}(X(t+1)=x+1|X(t)=x) = w_{x,t} \\
& \mathbb{P}(X(t+1)=x-1|X(t)=x) = 1- w_{x,t} 
\end{split}
\ee 

We will denote by $\langle X(t) \rangle_w$, or more simply $\langle X(t) \rangle$
the mean position in a given sample, and by $\overline{\cdots}$ the averages over samples. 
It is easy to see that the sample averaged bias and diffusion coefficient, defined as $\overline{\langle X(t) \rangle}= \bar v t$ and
$\overline{\langle X(t)^2 \rangle^c}= \bar D t$ are equal to
\be 
\bar v = 2 \langle w \rangle - 1 = \frac{\alpha-\beta}{\alpha + \beta}  \quad , \quad 
\bar D=\frac{4 \alpha \beta }{(\alpha + \beta)^2} 
\ee 
One can show that at large time the typical walk in a typical sample is also characterized
by the same bias and diffusion coefficient.
We will choose from now on $\alpha=\beta$, i.e $\bar v=0$ and $\bar D=1$. 

Since at large time the typical motion is diffusive \cite{BarraquandCorwinBeta} we will be 
interested in the following probability, which will be our observable
\be 
Z = Z_\xi(T)  = \mathbb{P}\left( \frac{X(T)}{ \sqrt{T/2}} > \xi \right) \,  , \label{defZxi} 
\ee 
where $\xi$ is an asymmetry parameter (here chosen positive $\xi \geq 0$)
which describes how the position of the random walker deviates from its
mean in a given sample. 
% \blue{where $\xi$ is a position parameter, describing the asymmetry
% of deviating from the average. Keep}
Keep in mind that $Z$ is a random variable with respect to the sample.
 We will thus be interested in the PDF's of $Z$ w.r.t. the sample, denoted $P(Z)$, equivalently of $H=\log Z$
denoted (abusively) as $P(H)$. At large time $T$ they
are expected to take the large-deviation forms
\bea 
&& P(Z) \sim e^{- \sqrt{T} \hat \Phi_\xi^{\rm RW}(Z)} \label{eq:rate:Z}\\
&& P(H) \sim e^{- \sqrt{T} \Phi_\xi^{\rm RW}(H)} 
\label{eq:rate:H}
\eea
and we will determine the rate functions $\hat \Phi_\xi^{\rm RW}(Z)$
and $\Phi_\xi^{\rm RW}(H)$, both through an analytical argument, and
through extensive numerics.

\section{Analytical predictions}
\label{sec:prediction}

\subsection{Continuum model observables}

One can define a similar observable for the continuum model 
\eqref{FP}-\eqref{Langevin}, namely
\be 
\tilde Z = \tilde Z_\xi(\tilde T)  = \int_{\xi \sqrt{\tilde T}}^{+\infty} \rmd y \, q_\eta(y,\tilde T) =
\mathbb{P}\left( \frac{y(\tilde T)}{\sqrt{\tilde T}} > \xi \right) \,  , \label{defZxicont} 
\ee 
where the particle is at the origin at time zero, $y(0)=0$. Note that here and below the tilde variables
are associated to the continuum model. We have shown in Ref.~\cite{KrajLedouCrossover} that at large time the PDF's of $\tilde Z$
and $\tilde H=\log \tilde Z$
take the large-deviation forms
\bea 
&& P(\tilde Z) \sim e^{- \sqrt{\tilde T} \hat \Phi_\xi(\tilde Z)} \label{eq:rate:Zcont}\\
&& P(\tilde H) \sim e^{- \sqrt{\tilde T} \Phi_\xi(\tilde H)} 
\label{eq:rate:Hcont}
\eea
and we have obtained the analytical expressions of the rate functions $\hat \Phi_\xi$ and
$\Phi_\xi$, which will be recalled below. These were obtained by considering
the following generating function, which takes the large-deviation form at large time
\be \label{generatingcont} 
\overline{ e^{- \tilde z \sqrt{\tilde T} \tilde Z} } \sim e^{- \sqrt{\tilde T}  \Psi_\xi(\tilde z) }
\ee 
The two rate functions are related by a Legendre transform. Indeed one has
\be 
\Psi_\xi(\tilde z) = \min_{\tilde Z \in [0,1]} ( z \tilde Z + \hat \Phi(\tilde Z) ) 
\ee 
In Ref.~\cite{KrajLedouCrossover} we obtained the expression of $\Psi_\xi(\tilde z)$
by two different methods, one of them will be recalled in the next section.
From it we obtained the rate functions $\hat \Phi_\xi$ and
$\Phi_\xi$ through Legendre inversion. The explicit formula for these
rate functions will be given in Section~\ref{subsec:explicit}. 

\subsection{Main prediction}

Our main prediction is that the discrete model is described by the same
rate functions as the continuum one, up to some scale factors,
hinting to a form of universality. More precisely 
we claim that 
\be 
\hat \Phi^{\rm RW}_{\xi}(Z) = \frac{\alpha}{\sqrt{2}} \hat \Phi_{\xi  }(Z) \quad , \quad
\Phi^{\rm RW}_{\xi}(H) = \frac{\alpha}{\sqrt{2}} \Phi_{\xi  }(H)  \label{prediction1} 
\ee 
and 
\be 
\Psi^{\rm RW}_{\xi}(z) = \frac{\alpha}{\sqrt{2}} \Psi_{\xi  }\left(\tilde z=\frac{z \sqrt{2} }{\alpha}\right) \label{prediction2}  
\ee 
We will now explain the origin of this prediction. 
To this aim we first need
to recall one method to obtain the rate function $\Psi_\xi$ in the continuum. 
Next we show how an extension of the same method for the discrete case leads to the predictions above. 

\subsection{Fredholm determinant method for the continuum model}

In \cite{BarraquandSticky} a mathematically well posed version of the continuum model,
called the sticky Brownian
motion, was defined. An exact formula was derived for the 
Laplace transform of the
PDF of $\tilde Z$ for any $\tilde T,\xi$ in terms of a (complicated) Fredholm determinant. For $u \geq 0$
one has  \cite[Thm~1.11]{BarraquandSticky}
\be \label{FD0} 
\overline{ e^{- u  \tilde Z_\xi(\tilde T) } } = \Det (I - K_u)\vert_{\mathbb{L}^2(C)} 
\ee 
where the kernel $K_u(v,v')$ acts on functions defined on a 
contour $C$ in the $v$ complex plane, where $C$ is a positively-oriented circle 
centered at ${\sf R}  $ with radius ${\sf R}$. The kernel reads
\be
\label{eq:KernelGuillaume0}
 K_u(v,v') = \frac{1}{2 \I \pi}
\int_{1/2 +\I \R} \frac{\pi u^s}{\sin \pi s}  \frac{g(v)}{g(v+s)}
\frac{\rmd s}{s+v-v'} 
\ee 
where the function $g(v)$ is  
\be
\label{eq:def_g}
g(v) = g_c(v) := e^{  \xi \sqrt{\tilde T} \psi_0(v) +  \tilde T \psi_1(v) } \Gamma(v) 
\ee 
where $\psi_{0,1}$ denote polygamma functions
(see \cite[Supp Mat Sec.~X]{KrajLedouCrossover} for more details and correspondence of conventions).
\\

In Ref.~\cite[Supp Mat Sec.~X]{KrajLedouCrossover} we have studied in detail the large time
limit $\tilde T \gg 1$ of the kernel $K_u$ 
\eqref{eq:KernelGuillaume0} and of the Fredholm determinant \eqref{FD0} 
when $u$ is scaled as $u = \tilde z \sqrt{\tilde T}$ for the continuum model (recalling that
here the observation time in the continuum model is denoted by $\tilde{T}$).  
From \eqref{generatingcont}  and \eqref{FD0} this provided an 
independent method to obtain the rate function $\Psi_\xi(\tilde z)$. 
The important point is that
we showed there that the only relevant quantity is the asymptotic form at large $\tilde T$
of the function $g(v)$ under the rescaling 
\be \label{scaling} 
v = w \sqrt{\tilde{T}} 
\ee 
%consistent with the scaling chosen in Eq.~\eqref{defZxicont}. 
More specifically one finds that this asymptotic form reads~\cite[Supp Mat Eqs.~(S206)-(S210)]{KrajLedouCrossover} 
\bea \label{logg} 
&& \log g_c(v) =
\sqrt{\tilde T} \left(\phi(w) + (w +  \xi) \log \sqrt{\tilde T}\right) \\ 
&&+ \chi(w) - \frac{1}{2} \log(\sqrt{\tilde T}) + o(\tilde T) \nn 
\eea
where we defined 
\be
\begin{split} \label{phiw}
\phi(w) &=   \frac{1}{w} - w + (w +  \xi) \log(w) \\
\chi(w) &= \frac{1}{2 w^2} - \frac{ \xi}{2 w} + \frac{1}{2} \log(2 \pi/ w)
\end{split}
\ee 
where we corrected a misprint in the last term in \cite{{KrajLedouCrossover}}. 
The knowledge of $\phi(w)$ in this asymptotic form then allows to obtain the explicit form 
of $\Psi_\xi(\tilde{z})$ in Eq.~\eqref{PsiFinal0}, see ~\cite[Supp Mat Sec.~X]{KrajLedouCrossover} 
for details. In particular, subdominant terms such as $\chi(w)$ are irrelevant.

\subsection{Discrete to continuum universality}

To obtain the rate functions for the discrete Beta random walk model, we can consider, similarly to 
\eqref{generatingcont}, the generating function associated to the observable $Z=Z_\xi(T)$ 
which takes the form at large time
\be \label{generatingRW}
\overline{ e^{- z \sqrt{T} Z} } \sim e^{- \sqrt{T}  \Psi^{\rm RW}_\xi(z) }
\ee 

One method to obtain $\Psi^{\rm RW}_\xi(z)$ is to use the result from
\cite{BarraquandCorwinBeta} which we now recall. 
Reference~\cite[Theorem 1.13]{BarraquandCorwinBeta} gives an exact formula for the Laplace transform \eqref{generatingRW} of the
PDF of $Z$ for any $T,\xi$ in terms of a (complicated) Fredholm determinant. For $u \geq 0$
one has
\be \label{FD} 
\overline{ e^{- u  Z_\xi(T) } } = \Det (I - K_u)\vert_{\mathbb{L}^2(C)} 
\ee 
where the kernel $K_u(v,v')$ acts on functions defined on a 
contour $C$ in the $v$ complex plane, where $C$ is a positively-oriented circle 
centered at ${\sf R} \geq 0 $ with radius ${\sf R}+\varepsilon$ so that $0 < \varepsilon \leq\min(1,\alpha+\beta)$. 
The kernel has the same form as for the continuum case, namely
\be
\label{eq:KernelGuillaume}
 K_u(v,v') = \frac{1}{2 \I \pi}
\int_{1/2 +\I \R} \frac{\pi u^s}{\sin \pi s}  \frac{g(v)}{g(v+s)}
\frac{\rmd s}{s+v-v'} 
\ee 
except that now for the Beta random walk one has
\be
\begin{split}
& g(v)=g_{\rm RW}(v) \label{gRW} \\
& = \left(\frac{ \Gamma(v)} {\Gamma(\alpha + v) }\right)^{\frac{T- \xi \sqrt{\frac{T}{2}}}{2}} 
\!\! \left(\frac{ \Gamma(\alpha + \beta + v)} {\Gamma(\alpha + v) }\right)^{\frac{T+ \xi \sqrt{\frac{T}{2}}}{2}} \Gamma(v) 
\end{split}
\ee 
where we recall that from now on we restrict to the case $\beta=\alpha$.

%In \cite{BarraquandSticky} a mathematically well posed version of the continuum model,
%called the sticky Brownian
%motions, was defined and it was shown that the same formula \eqref{FD} 
%for $\overline{ \exp(- u  \tilde Z ) }$ holds, with the same contour with the choice 
%$\varepsilon=0$, the only difference being that 
%the function $g(v)$ is now 
%\be
%\label{eq:def_g}
%g(v) = g_c(v) = e^{  \xi \sqrt{\tilde T} \psi_0(v) +  \tilde T \psi_1(v) } \Gamma(v) 
%\ee 
%where $\psi_{0,1}$ denote polygamma functions
%(see \cite[section X]{KrajLedouCrossoverSM} for more details and correspondence of conventions).
%\\ 

It is interesting to note that there is a way to obtain the continuum model from the discrete one, by taking
the limit $\alpha \to 0$. Indeed if one sets 
\be
T=2 \tilde T/\alpha^2  \label{correspondence} 
\ee
one has,
with the same value of $\xi$ 
\be 
\lim_{\alpha \to 0} g_{\rm RW}(v)  = g_c(v) 
%\Gamma(v) \exp \left( \hat Y \psi_0(v) + \frac{\hat T}{2} \psi_1(v)  \right)   
\ee 
This corresponds to the convergence of the discrete random walk to the continuum one,
which can be expressed as the convergence \cite{BarraquandSticky} 
\be 
\alpha X(2 \alpha^{-2} \tau) \underset{\alpha \to 0}{\to} y(\tau) 
\ee 
recalling that $X(t)$ corresponds to position in the Beta random walk with index $\alpha$,
and $y(\tau)$ to the position of the particle in the continuum model \eqref{Langevin}.

However this is not what we are interested in here. Instead we want to keep $\alpha$ fixed and
take the time $T$ of the Beta random walk to be large. We now argue that it leads to the
same large-deviation rate functions as for the continuum model, up to the rescaling \eqref{correspondence}. \\

%In Ref.~\cite[section X]{KrajLedouCrossoverSM} we have studied in detail the large time
%limit $\tilde T \gg 1$ of the Fredholm determinant \eqref{FD} and of the kernel $K_u$ 
%\eqref{eq:KernelGuillaume} for the continuum model (recalling that
%here the continuous time is denoted by $\tilde{T}$). We will thus heavily rely on that
%study and not reproduce here the various steps. The important point is that
%we showed there that the only relevant quantity is the asymptotic form at large $\tilde T$
%of the function $g(v)$ under the rescaling 
%\be \label{scaling} 
%v = w \sqrt{\tilde{T}} 
%\ee 
%%consistent with the scaling chosen in Eq.~\eqref{defZxicont}. 
%More specifically one finds that this asymptotic form reads~\cite[Eqs.~(S206)-(S210)]{KrajLedouCrossoverSM} 
%\bea \label{logg} 
%&& \log g_c(v) =
%\sqrt{\tilde T} \left(\phi(w) + (w +  \xi) \log \sqrt{\tilde T}\right) \\ 
%&&+ \chi(w) - \frac{1}{2} \log(\sqrt{\tilde T}) + o(\tilde T) \nn 
%\eea
%where we defined 
%\be
%\begin{split}
%\phi(w) &=   \frac{1}{w} - w + (w +  \xi) \log(w) \\
%\chi(w) &= \frac{1}{2 w^2} - \frac{ \xi}{2 w} + \frac{1}{2} \log(2 \pi/ w)
%\end{split}
%\ee 
%where we corrected a misprint in the last term in \cite{{KrajLedouCrossoverSM}}. 

Since the form of the kernel is quite similar in both cases, 
to obtain the asymptotics of $K_u$ in \eqref{eq:KernelGuillaume} and of the 
Fredholm determinant \eqref{FD} for the Beta random walk, in the limit 
$T \to +\infty$ with $u=z \sqrt{T}$, we also only need to study
the large time limit of the function $g_{RW}(v)$ under the same rescaling \eqref{scaling}.
Although we are working here for an arbitrary fixed $\alpha$, we will choose 
the correspondence between the discrete and continuous time as in
\eqref{correspondence}. Let us use the 
expansion at large $v$ 
\begin{widetext}
\be
v^{b-a} \frac{\Gamma(a+v)}{\Gamma(b+v)} = 1+\frac{(a-b)
   (a+b-1)}{2 v}  +\frac{(a-b-1) (a-b)
   \left(3 b^2+6 a b-5 b+a (3a-7)+2\right)}{24v^2} + \mathcal{O}\left(\frac{1}{v^3}\right) 
\ee
\end{widetext}
for any $a,b=\mathcal{O}(1)$. Let us consider
\eqref{gRW} with $\beta=\alpha$, express it as a function of $\tilde T$ using
\eqref{correspondence},
and insert the rescaling \eqref{scaling}. In the large $\tilde T$ limit one
finds
\be 
\begin{split}
& \log g_{RW}(v) = 
\sqrt{\tilde T} \left(\phi(w) + (w +  \xi) \log \sqrt{\tilde T}\right) \\ 
&+ \chi_{RW}(w) - \frac{1}{2} \log(\sqrt{\tilde T}) + o(\tilde T) 
\end{split}
\ee
with  
\be 
\chi_{RW}(w)  = (1 - 2 \alpha) ( \frac{1}{2 w^2}  - \frac{\xi}{2 w} ) + \frac{1}{2} \log(2 \pi/w)
\ee 
and the function $\phi(w)$ being identical to the one for the continuum model in \eqref{phiw}.

Thus, in the large time limit we can identify $Z=\tilde Z$, i.e. the two random variables
\be 
Z_\xi(T) \equiv \tilde Z_\xi(\tilde T)
\ee 
and identify separately each sides of \eqref{generatingRW} and \eqref{generatingcont} respectively
which leads to 
\bea 
&& z \sqrt{T} \equiv \tilde z \sqrt{\tilde T} \\
&& \sqrt{T} \Psi^{\rm RW}_\xi(z) = \sqrt{\tilde T} \Psi_\xi(\tilde z) 
\eea 
which using the correspondence between continuum and discrete time in \eqref{correspondence}, finally leads to the prediction \eqref{prediction1} and \eqref{prediction2}
for the rate function of the Beta random walk.

% \kraj{
% \begin{equation}
%     \chi_{RW}(w)= \frac{(-1+2 \alpha)(-1 + w \xi) }{ 2w^2} 
% - \frac{1}{2} \log(\frac{w}{2\pi})
% \end{equation}
% }

\subsection{Explicit formula for the rate functions} 
\label{subsec:explicit} 

We now recall the analytical prediction from \cite{KrajLedouCrossover}
for the rate functions of the continuum model. 
Since $Z=\tilde Z$ and $H=\tilde H$, see section above, we use below
only the notations $Z$ and $H$ in place of $\tilde Z$ and $\tilde H$.
The rate function $\hat \Phi_\xi(Z)$ is obtained from the parametric representation 
\be 
\begin{cases}
\hat \Phi_\xi(Z)= \Psi_\xi(z)- \tilde z Z ,\\
 Z = \Psi_\xi'(\tilde z)\, .
\end{cases}
\label{eq:ParametricRepresentationZ}
\ee 
where $\Psi_\xi(\tilde z)$ for $\xi \geq 0$ is given by 
\be \label{PsiFinal00} 
\begin{split}
\Psi_\xi(\tilde z) &= -
\dashint_\R \frac{\rmd q}{2 \pi}\frac{\mathrm{Li}_2( \tilde z (\I q - \frac{\xi}{2}) e^{-q^2 - \frac{\xi^2}{4}} )}{(\I q - \frac{\xi}{2})^2} %+z\Theta(-\xi)
\\
\end{split}
\ee 
where the principal value is required only for $\xi=0$. 

We now consider here only the
case $\xi=0$ where for any real value of $\tilde z$, the dilogarithm in the integrand of \eqref{PsiFinal00} 
does not have any branch cut on the real axis for $q$. 

This expression for the rate function $\Psi_{\xi=0}(\tilde z)$ then allows 
to obtain $\hat \Phi_{\xi=0}(Z)$ for any $Z \in [0,1]$. From this
one obtains the 
rate function $\Phi_{\xi=0}(H)=\hat \Phi_{\xi=0}(Z)$ by the simple change of
variable $H=\log Z$ for any $H\leq 0$. This is summarized in the Table~\ref{tab:app-case1}.

\begin{table}[h!]
    \centering
    \begin{tabular}{p{2cm}  p{1.9cm}  p{1.4cm}   p{2.2cm} }
    \hline 
        \hline \\[-0.8ex]
        interval of $H$ & interval of $\tilde z$ & $H=$ & $\Phi_\xi(H)=$ \\[0.5ex]
        \hline &&&\\[-0.5ex]
        $H \in \R^-$ & $\tilde{z} \in \R$ &$\log \Psi_0'(\tilde z)$ & $\Psi_0(\tilde z)- \tilde z \Psi_0'(\tilde z)$\\[1ex]
        \hline 
        \hline
    \end{tabular}
    \caption{Case $\xi=0$}
    \label{tab:app-case1}
\end{table}

The case $\xi>0$ is more
involved and is given in the Appendix~\ref{sec:supp-mat-table-legendre}.
However one can give for any $\xi$ the typical value 
\bea \label{Ztyp} 
 Z_{\rm typ}=\overline{Z} = \Psi_\xi'(0) = \frac{1}{2} {\rm Erfc}\left(\frac{\xi}{2}\right)\nn
\eea 
and the variances of the PDF's $P(Z)$ and $P(H)$ for the continuum model \cite{KrajLedouCrossover}
\bea 
\label{eq:variances-Z-H}
&& \overline{Z^2}^c = \frac{1}{4 \sqrt{2 \pi \tilde T} } e^{-\xi^2/2} \\
&& \overline{H^2}^c = \frac{1}{\sqrt{2 \pi \tilde T}} e^{-\xi^2/2} ({\rm Erfc(\frac{\xi}{2}))^{-2}}
\eea   
The corresponding variances for the Beta random walk at large time are obtained by the correspondence
$\tilde T= \frac{\alpha^2}{2} T$.

\section{Methods}
\label{sec:methods}

Next, we describe our numerical approaches. In Subsection~\ref{subsec:def-random-walk-numerics}, we first state how we obtain, for each given sample $\omega=\{w_{x,t}\}$, 
as drawn from the beta distribution \eqref{eq:beta:distr}, the quantities $Z$ according to \eqref{defZxi} and therefore $H=\log Z$. We are interested in the distributions $P(Z)$ and $P(H)$. In Subsection~\ref{subsec:intro-importance} we explain how we achieve this over a large range of the support down to very small probability densities such as $10^{-50}$ or even smaller.

\subsection{Random walk on a lattice}
\label{subsec:def-random-walk-numerics}
For each of the samples $\omega=\{w_{x,t}\}$, corresponding to the probabilities \eqref{eq:move:probs} to move left and right, we calculate the probability $Q(X|t)$ of reaching site $X$ at step $t$. For this purpose we apply a dynamic programming, i.e., transfer matrix, approach by calculating
\begin{eqnarray}
 Q(0|0) & = & 1 \nonumber\\
 Q(X|0) & = & 0 \; \text{for} \; X\neq 0 \\
 Q(X|t+1) & = & w_{X-1,t} Q(X-1|t) + \nonumber \\
 & & (1-w_{X+1,t})Q(X+1|t)\; \nonumber
\end{eqnarray}
for $t=0,1,\ldots, T-1$.
For a walk of $T$ steps, these probabilities can be calculated in $\mathcal{O}(T^2)$ time. 
% \akh{I did not use tricks to optimize the calculation. For $\xi=0$ one
%  could save 50\% of the calculation of the half matrix, for $\xi>0$ even less ... not worth adapting 35 pages of code (MPI part etc included) to that.}
This allows one to obtain the cumulative probability $R$ of being right of some point $X$ by simply summing 
\be
R(X|T) = \sum_{X' > X} Q(X'|T)\,,
\ee
which is achieved in $\mathcal{O}(T)$ steps, which is negligible compared to the $\mathcal{O}(T^2)$ steps to compute the (half) transfer matrix $Q(X|t)$. The value $Z$ of \eqref{defZxi} we are interest in is obtained by 
\be
\label{eq:Z}
Z=Q(\sqrt{T/2} \, \xi \,| \,T)\,
\ee
where we round $\sqrt{T/2}\, \xi$ to the next lowest integer.
Note that for small values of $\xi$, not all values of the matrix $Q(X|t)$ contribute. But even for $\xi=0$, where
walks contribute which reach $X=T/2$ and return to $X=0$,
one needs half of $Q$. Thus, the total computation time
is always $\mathcal{O}(T^2)$. 

The corresponding value 
of $H$ is obtained simply by $H=\log(Z)$. Note that $H$ is completely
determined by the sample $\omega$, so we can write $H=H(\omega)$.

\subsection{Introduction to importance sampling}
\label{subsec:intro-importance}
For the purpose of the introduction of the idea of importance sampling, we retain some elements of the presentation made in Ref.~\cite{NumericsHartmann}.  In principle one could obtain an estimate of the probability distribution
$P(H)$ numerically from \textit{direct
sampling}. For this, one generates many disorder {samples} and calculates  $H=\log Z$ for each one according to Eq.~\eqref{eq:Z}.
Then  the
distribution is estimated by the suitably normalized histogram  of the values of $H$. Nevertheless, this limits the smallest probabilities which can be resolved to the inverse of the number of samples, hence reaching probabilities as small as $10^{-50}$ is strictly impossible. Therefore, a different approach is required.

To estimate $P(H)$ for a much larger range, where probability densities as small as $10^{-50}$ may appear, we use a more powerful approach, called  importance sampling as discussed in Refs.~\cite{align2002,largest-2011}.  This approach has been successfully applied to many problems in statistical physics and mathematics 
to obtain the tails of distributions arising in equilibrium and non-equilibrium situations \cite{rare-graphs2004,partition2005,monthus2006,rnaFreeDistr2010,driscoll2007,saito2010,fBm_MC2013,work_ising2014,convex_hull2015,convex_hull_multiple2016}. The idea behind importance sampling is to sample the different disorder samples with
an additional bias $\exp(-\theta H(\omega))$ where $\theta$  is an adjustable parameter interpreted as a fictive temperature. If $\theta>0$ the samples with a negative $H$ become more likely, conversely
if $\theta<0$ the samples with a positive $H$ are favored.   
Now, it is not possible to sample the disorder samples $\omega$ directly when the bias is included. For this reason,
a standard Markov-chain Monte Carlo simulation is used to sample according to the biased distribution 
 \cite{newman1999,landau2000}. Here, one has a disorder sample $\omega$ as current configuration of the Markov chain, and the configurations change only slightly from step to step. In detail,
 at each step of the Markov chain, a new disorder sample $\omega^*$ is proposed by replacing on the current sample $\omega$ 
a certain fraction $r$ of the random numbers $\omega=\{w_{x,t}\}$ by new random numbers which are drawn according to Eq.~\eqref{eq:beta:distr}.
The new disorder sample is then accepted with the usual Metropolis-Hastings probability
\begin{equation}
p_{\rm Met} = \min\lbrace1,e^{-\theta\left[H(\omega^*)-H(\omega)\right]}\rbrace,
\end{equation}
otherwise the old configuration is kept \cite{metropolis1953}.
 By construction, the algorithm fulfils detailed balance and is ergodic, since within a sufficient number of steps, each possible sample may be constructed. Thus,
in the limit of infinitely long Markov chains, the distribution of biased disorder samples will follow the probability
\begin{equation}
q_\theta(\omega) = \frac{1}{W(\theta)} P_{\text{dis}}(\omega)e^{-\theta H(\omega)}\,, \label{eq:qT}
\end{equation}
where $ P_{\text{dis}}(\omega)$ is the original disorder distribution, i.e., the product of the Beta distributions for all disorder values, and $W(\theta)= \sum_{\omega} P_{\text{dis}}(\omega)e^{-\theta H(\omega)} $ is the normalisation factor. Note that $W(\theta)$ also depends on the walk length
$T$ because of finite-size effects. $W(\theta)$ is generally unknown but can be determined, see below. The output of this Markov chain allows one to construct a biased histogram $ P_\theta(H)$. In order to get the correct empirical probability density $P(H)$ one should unbias the result such that
\begin{equation}
 P(H) =  e^{\theta H} W(\theta) P_\theta(H).
\label{eq:rescaling}
\end{equation}
Hence, the target distribution $P(H)$ can be estimated, up to a normalisation
constant $W(\theta)$. For each value of the
parameter $\theta$, 
a specific range of the distribution $P(H)$ will be sampled and  using a positive (respectively, negative) parameter allows one to sample the region of a distribution at the left (respectively, at the right) of its center. 

For suitably chosen sets of temperature values $\theta$, the ranges of support for neighboring densities $P_{\theta_i}(H)$ and $P_{\theta_{i+1}}(H)$
will overlap. Since after rescaling with $W(\theta_i)$ and $W(\theta_{i+1})$, respectively, they must be equal to $P(H)$. Thus, in particular they have to be in equal to each other, up to statistical fluctuations, for those values of $H$ where they overlap. This allows one to determine rations $W(\theta_i)/W(\theta_{i+1})$ for all neighboring pairs of temperatures, and finally all absolute values $W(\theta_i)$ through the overall normalisation of $P(H)$, for details and examples see Appendix~\ref{app:sec:technical-details} and Refs.~\cite{align2002,largest-2011}. Most accurately, the determination of the normalisation factors can be achieved using the Multi Histogram approach \cite{ferrenberg1989}, see also the convenient tool of Peter Werner \cite{werner2022gluer}.

\section{Comparison of the theoretical predictions with the simulations}
\label{sec:compare}

We now compare the theoretical predictions of Section \ref{sec:prediction} with the numerical simulations of the finite-time random walks on a lattice, for various values of $\xi$.
We insist on the fact that the comparison will be done without any fitting parameter.

\subsection{Presentation of the simulations}

 The numerical simulations were run for walks of length $T\in \lbrace  64, 128, 256, 512, 1024 \rbrace$, the largest lengths only for some cases. Most of the walks are for distribution parameter $\alpha=1$, which corresponds
 to a uniform $U(0,1)$ distribution, but in the beginning we also show some simple sampling results for other values of $\alpha$, which indicate the universality with respect to $\alpha$ subject to simple scaling of the number of steps. We have evaluated the cumulative distribution
 of positions for several  values of the asymmetry %position 
 parameter $\xi=\{0,1,2,3,4,5\}$.

\begin{figure}[!ht]
    %\centering
    \includegraphics[width=0.9\linewidth]{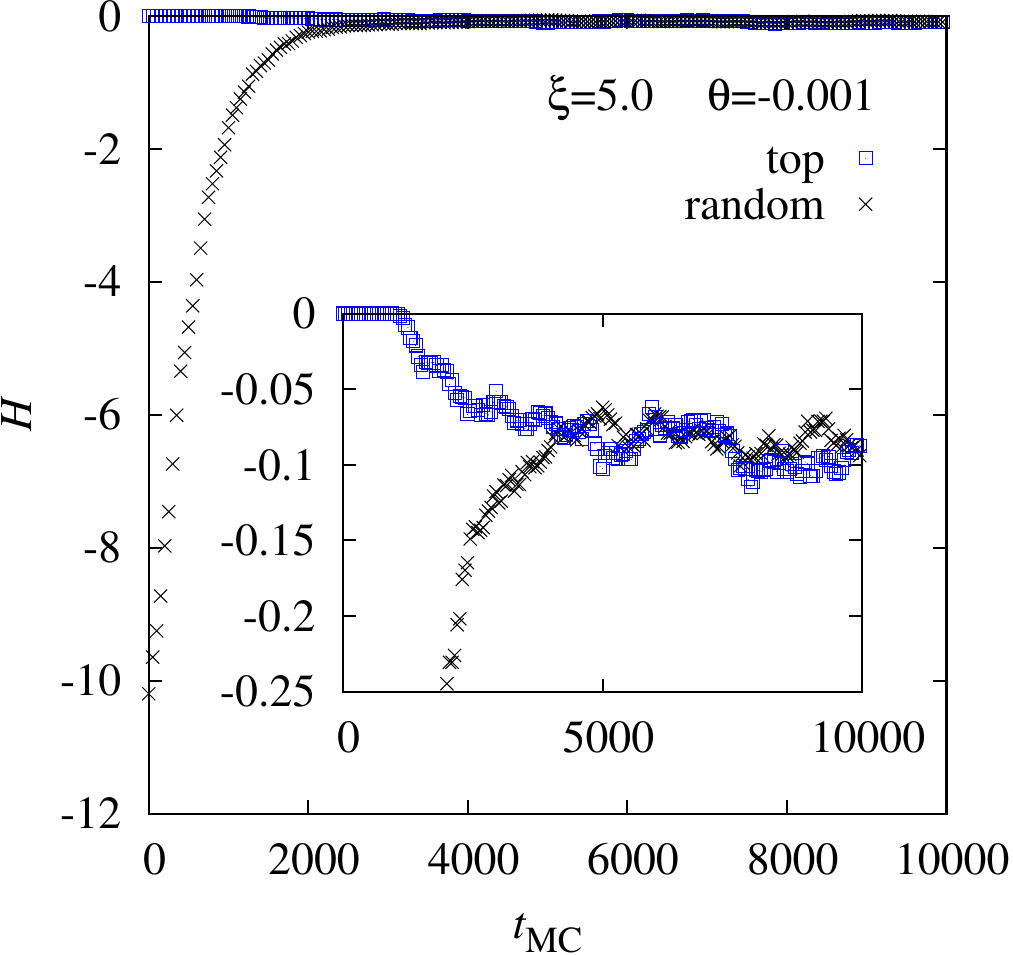}
    \caption{Equilibration of the Markov chain: Measured value of $H$ as function of the number $t_{\mathrm{MC}}$ of Markov steps for the case $\xi=5$, $T=128$ and sampling temperature $\Theta=-0.001$ which corresponds to the very tail of the distribution which is hardest to reach. Two initial starting configurations $\omega$ resulting in very different initial values of $H$ were chosen: one just typical random one, and one
    where $w_{x,t}=0.999$ for all values ("top"). The inset enlarges the top part.}
    \label{fig:equilibration}
\end{figure}

 For the large-deviation simulations, we have to make sure that the Markov chain is equilibrated. This can be confirmed by running the Markov chain for very different initial configurations of the sample $\omega$. Extreme samples where all entries $\omega_{x,t}$ are close to 0, or all values are close to 1, correspond to extreme values of $H$. An impression of the convergence of the
 Markov chain is obtained monitoring $H(t_{\mathrm{MC}})$ as a function of the number $t_{\mathrm{MC}}$ of Monte Carlo steps and observing where these values agree within fluctuations for different initial configurations of  $\omega$, see Fig.~\ref{fig:equilibration}. Evidently, the equilibration is obtained rather quickly, within few thousand MC steps.

  \subsection{Variance for the case $\xi=0$}

\begin{figure}[htb]
    \centering
    \includegraphics[width=0.9\linewidth]{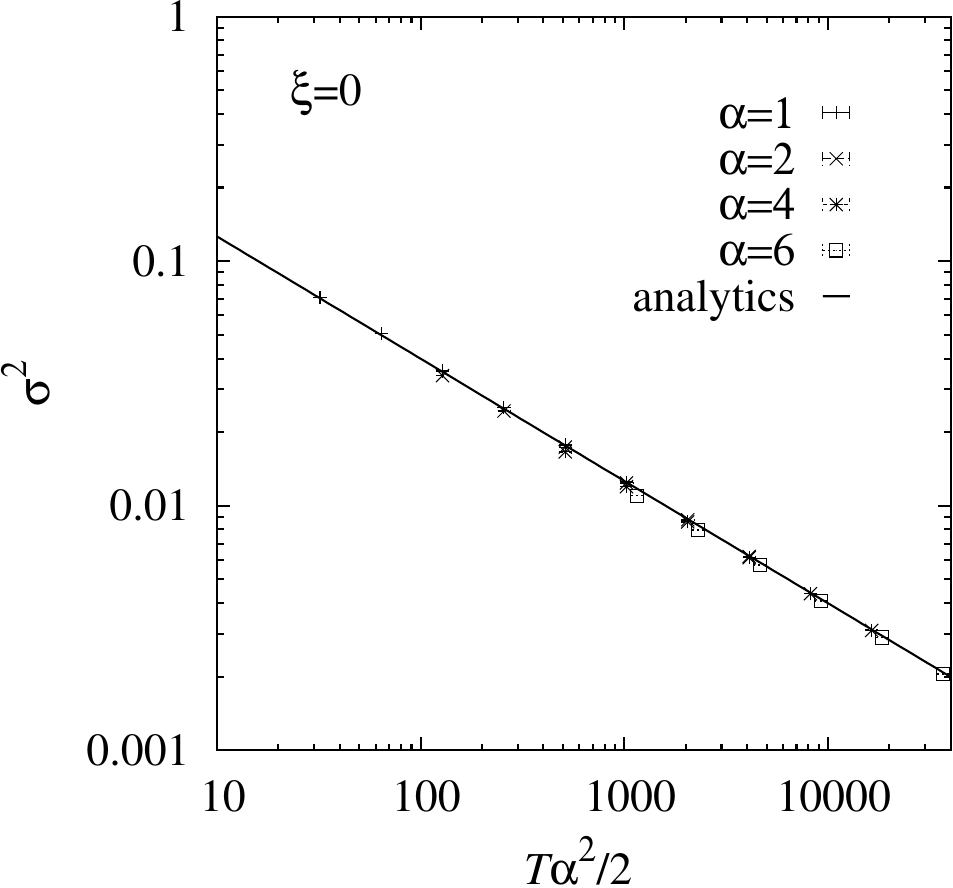} \hspace*{0.05\textwidth}
 \caption{Variance $\sigma^2$ of the distribution $P(H)$ for different values of $\alpha$,
 as a function of the scaled parameter $\tilde T= \frac{\alpha^2 T}{2}$
 where $T$ is the number of steps of the Beta random walk. It is compared to the analytical prediction \eqref{predictionsigma} (solid line).
 }
 \label{fig:variance}
 \end{figure}

First, we consider the probability distribution $P(H)$
as obtained by simple sampling
for $\xi=0$ and
several values of the distribution parameter $\alpha$. We determine its variance $\sigma^2= \overline{H^2}^c$ as a function of %the final time
the total length of the random walk
$T$, up to $T=2048$.
Our analytical prediction at large $T$ is, see \eqref{eq:variances-Z-H}
\be \label{predictionsigma} 
\sigma^2 \simeq \frac{1}{\sqrt{2 \pi \tilde T}} \simeq \frac{1}{\sqrt{2 \pi}} \sqrt{\frac{2}{\alpha^2 T}} 
\ee 
In Fig.~\ref{fig:variance} we show the variance $\sigma^2$ as a function of the scaled time parameter $\tilde T= \frac{\alpha^2 T}{2}$.
As visible, the data points fall nicely on one line, proving the universality with respect to $\alpha$.

\begin{figure}[htb]
 \centering
    \includegraphics[width=0.9\linewidth]{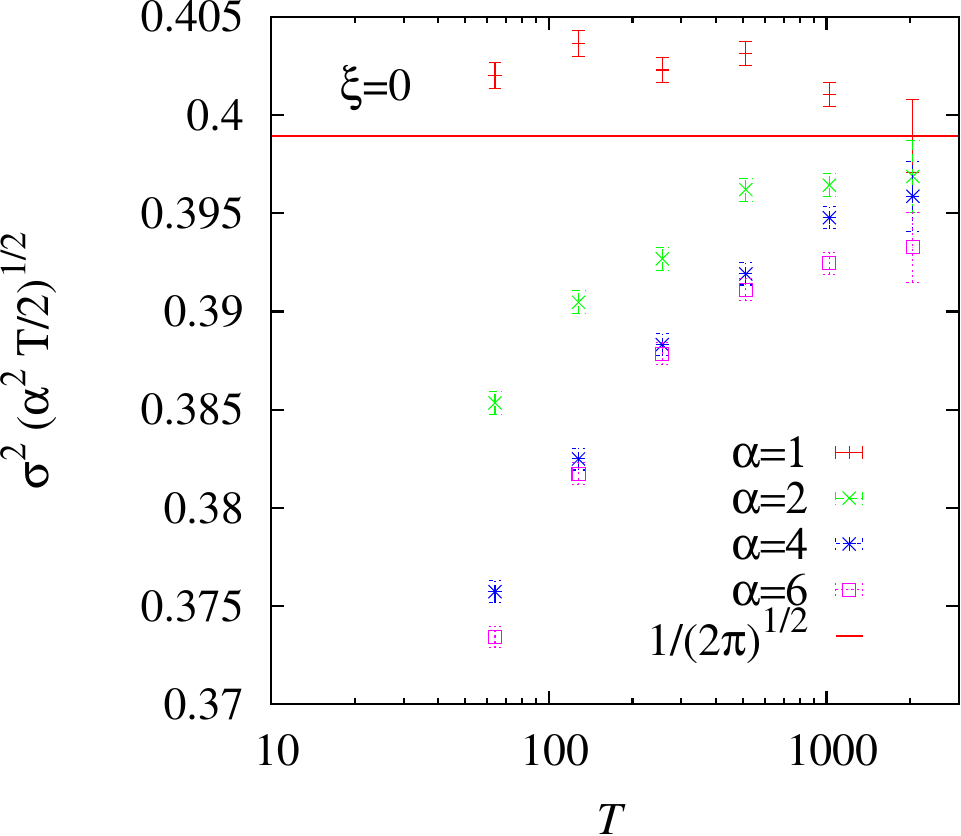}
   \caption{ The variance $\sigma^2$ scaled by the expected limiting behavior, as function of step size $T$. The prediction $1/\sqrt{2 \pi}$ from \eqref{predictionsigma} is also shown. 
    \label{fig:variance_scale}}
\end{figure}

To see how well the expected limiting behavior \eqref{predictionsigma}  
is reached, we plot in Fig.~\ref{fig:variance_scale} the combination $\sigma^2 \tilde T^{1/2}=\sigma^2 (\alpha^2 T/2)^{1/2}$.
For all considered values of $\alpha$, a convergence to the expected value $1/(2\pi)^{1/2}$ is visible. The convergence seems to be faster for smaller values of $\alpha$, i.e., for more flat step distributions of the samples  $\omega$.

\subsection{Distribution $P(H)$ }

\begin{figure}[!htb]
    \centering
    \includegraphics[width=0.9\linewidth]{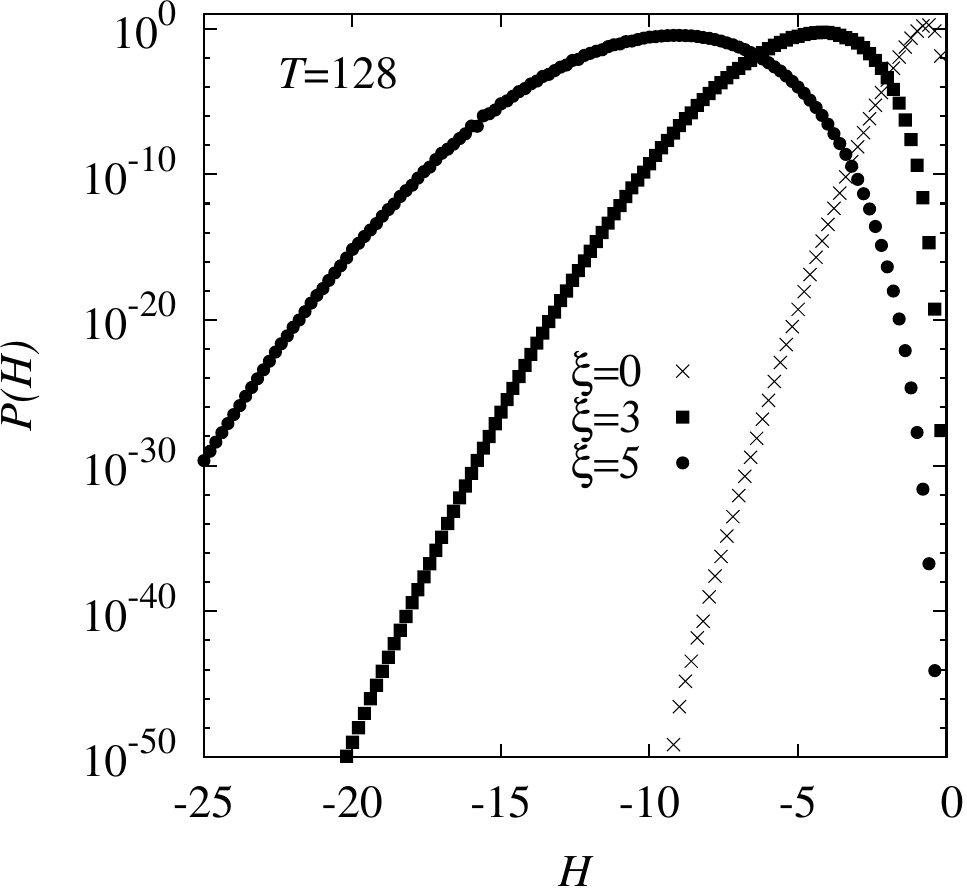}
    \caption{Distribution $P(H)$ for walk length $T=128$ and three values of the asymmetry parameter $\xi$.}
    \label{fig:distrH}
\end{figure}

The distribution of $H$ is shown in Fig.~\ref{fig:distrH} for $T=128$ and three values of the asymmetry parameter $\xi$. As visible, with the large-deviation approach, here small probabilities such as $10^{-50}$ are reached. For increasing values of $\xi$, the probability of a walk ending beyond $\xi\sqrt{T/2}$ will decrease, which is reflected by a shift of the distribution $P(H)$ to more negative values of $H$. For a more detailed analysis and comparison with the analytic results, we consider from now on the rate functions.

\subsection{Rate functions}

\begin{figure*}[!htb]
    \centering
    \includegraphics[width=0.45\linewidth]{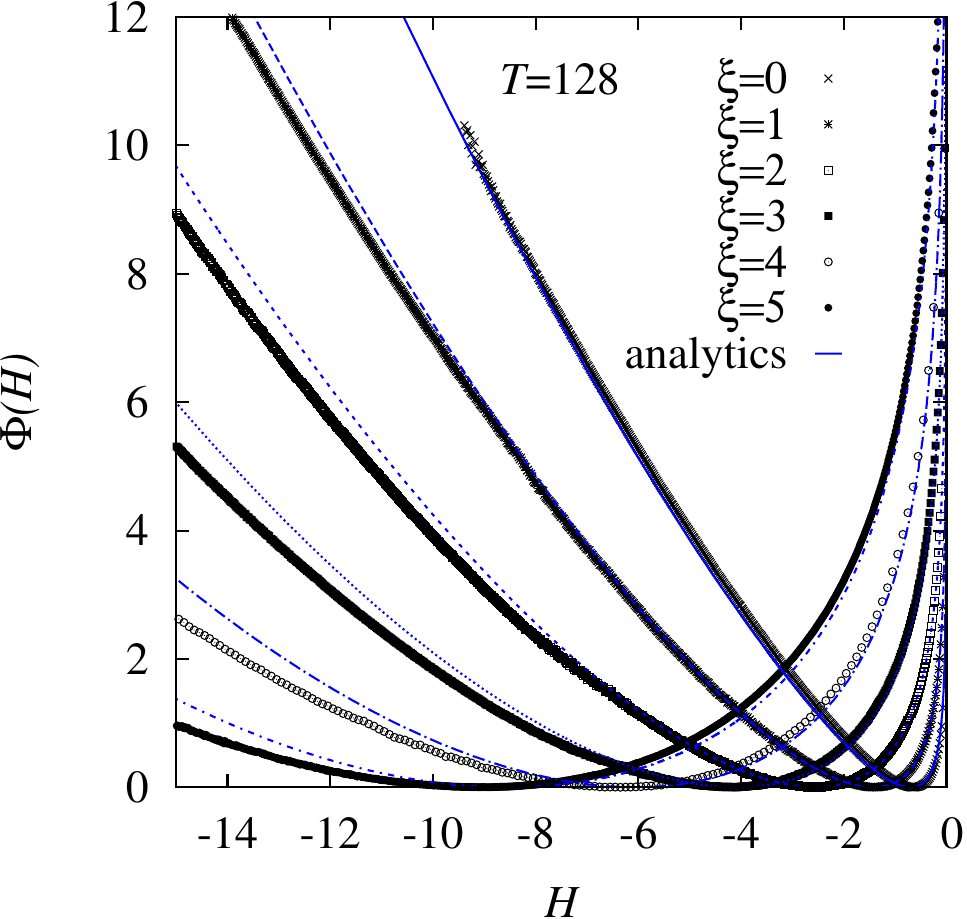}
    \hfill
    \includegraphics[width=0.45\linewidth]{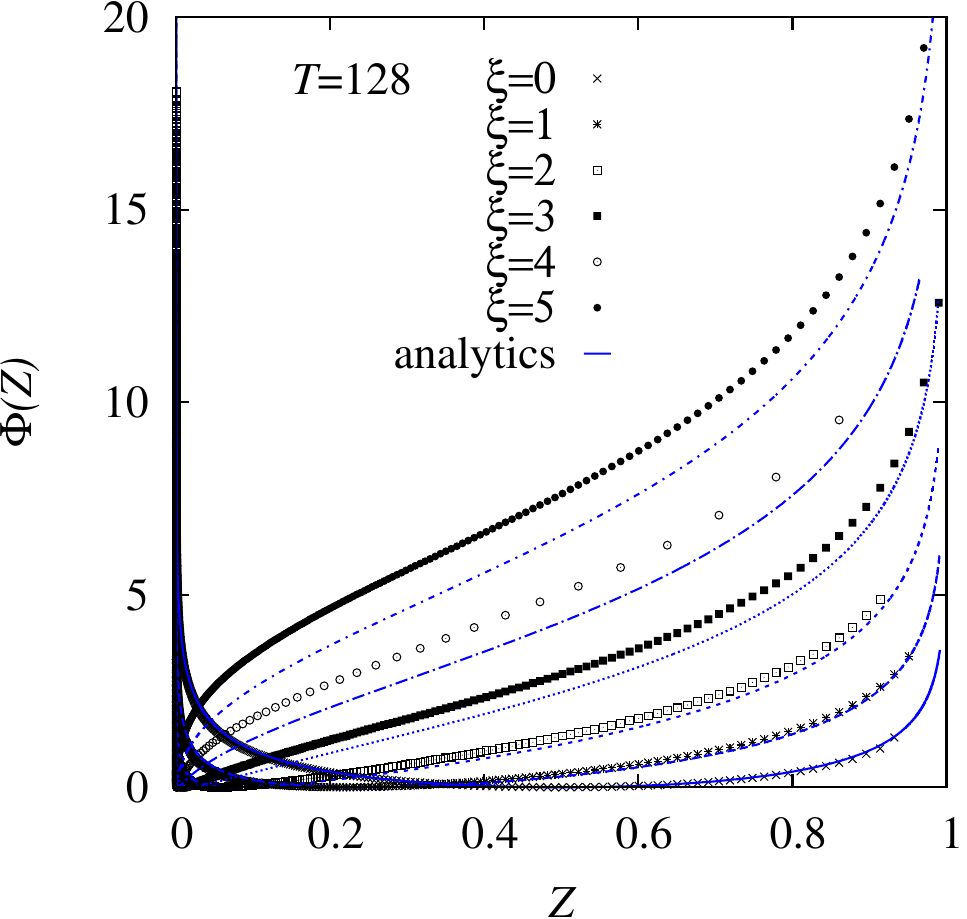}
    \caption{Rate functions $\Phi(H)$ and $\Phi(Z)$ for walk length $T=128$ and various values of the asymmetry parameter $\xi=\{0,1,2,3,4,5 \}$. The lines show our analytical predictions for $T\to\infty$.}
    \label{fig:rate_fcts}
\end{figure*}

We will now test the analytical prediction \eqref{prediction1} for the rate functions 
$\Phi^{\rm RW}_\xi(H)$ and $\hat \Phi^{\rm RW}_\xi(Z)$ defined in Eqs.~\eqref{eq:rate:Z} and \eqref{eq:rate:H}
(where the rate functions on the r.h.s. of \eqref{prediction1} are given in Sections \ref{subsec:explicit} and Appendix \ref{app:analytical}). 
Note that in this numerical section we simplify notations and denote
$\Phi(H) \equiv \Phi^{\rm RW}_\xi(H)$ and $\Phi(Z) \equiv \hat \Phi^{\rm RW}_\xi(Z)$. 
These rate functions are shown in Fig.~\ref{fig:rate_fcts} for walk length $T=128$ and all considered values of the asymmetry parameter $\xi= \{0,1,2,3,4,5 \}$. Note that a value of $\Phi$ close to 20 correspond for $T=128$ to a probability
$e^{-\sqrt{128}\times 20}\approx 5\times 10^{-99}$.

% We perform a comparison with the analytical prediction \eqref{} via the rate functions $\Phi(H)$ and $\Phi(Z)$ as defined in Eqs.~\eqref{eq:rate:Z} and \eqref{eq:rate:H}.  

\begin{figure*}[!htb]
    \centering
    \includegraphics[width=0.45\linewidth]{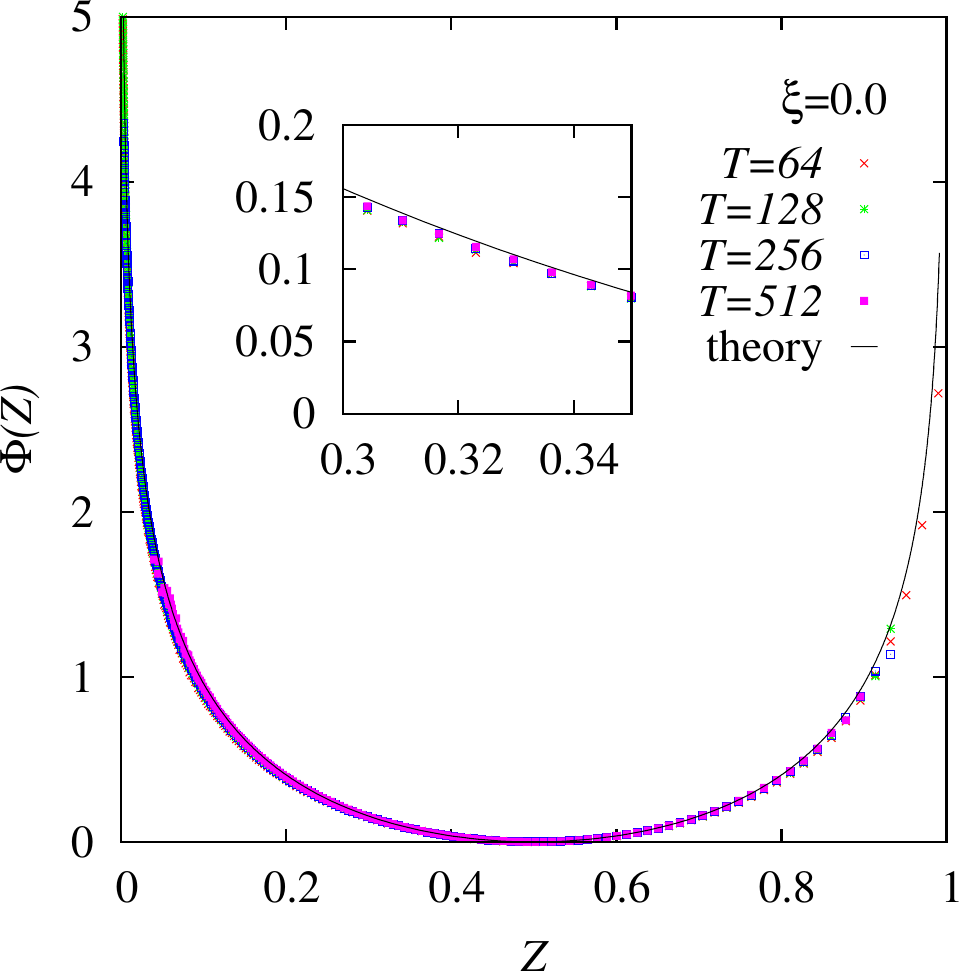}
    \hfill
        \includegraphics[width=0.45\linewidth]{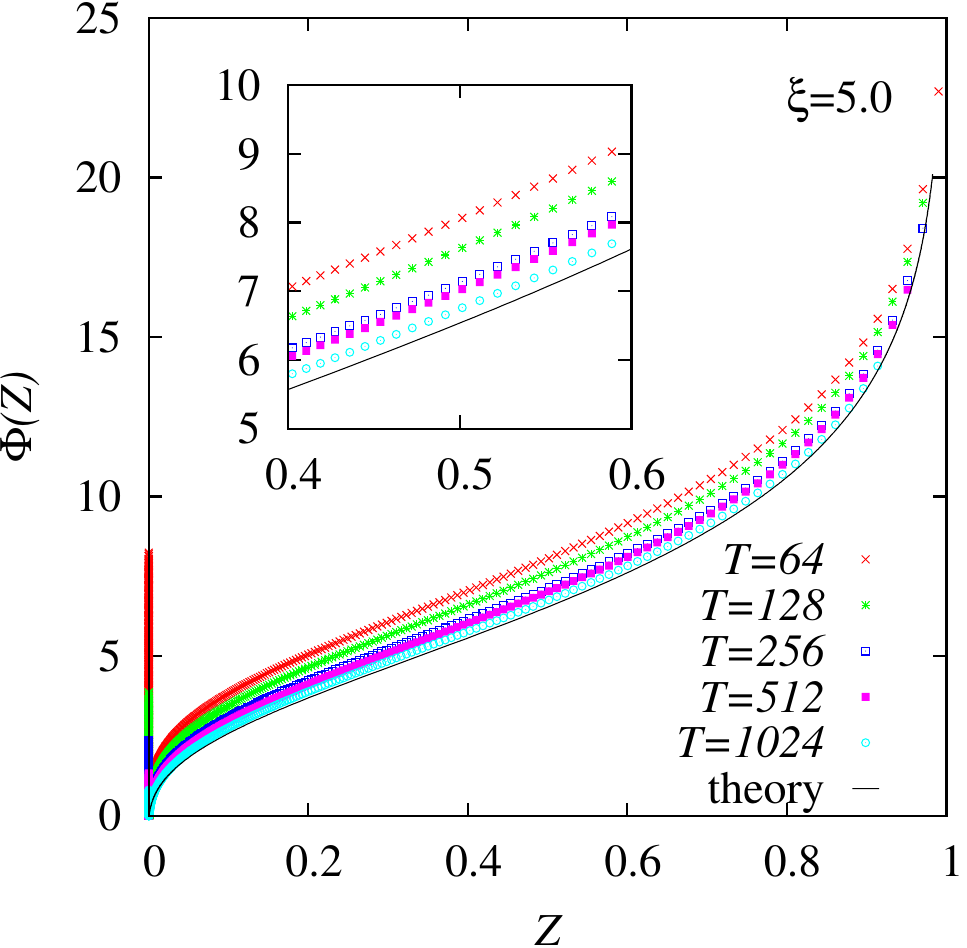}
    \caption{Rate function $\Phi(Z)$ for various walk lengths $T=\{64,128,256, 512 \}$ and 1024, for the cases $\xi=0$ (left, only up to $T=512$) and $\xi=5$ (right). The lines show the analytical results, respectively. The insets enlarge the regions $I\in[0.3,0.35]$ (left) and $Z\in[0.4,0.6]$ (right).
    \label{fig:scaling:rate_fct}}
\end{figure*}

For small values of $\xi$ already a good agreement between finite-$T$ numerical data and analytical results is visible. Nevertheless, for values such as $\xi=4$ and $\xi=5$ substantial deviations are visible. For this reason, we have performed numerical simulations for the two extreme cases of the asymmetry $\xi=0$ and $\xi=5$ for various lengths of the walk $T=\{ 64,128,256,512 \}$, and even $T=1024$ for $\xi=5$. The results for $\Phi(Z)$ are shown in Fig.~\ref{fig:scaling:rate_fct}. For the case $\xi=0$ basically all results agree, the limiting behavior is already visible for short walk length $T$. For $\xi=5$ a clear convergence to the analytical result is visible. The fact that the finite-$T$ corrections are stronger for larger values of $\xi$ reminds one of the different convergence speeds within the central limit theorem: The properly rescaled sum of random numbers attains a Gaussian shape near the typical values, corresponding to small values of $\xi$ here, much faster than in the tails, corresponding to large values of $\xi$. 

\begin{figure}[!htb]
    \centering
    \includegraphics[width=0.9\linewidth]{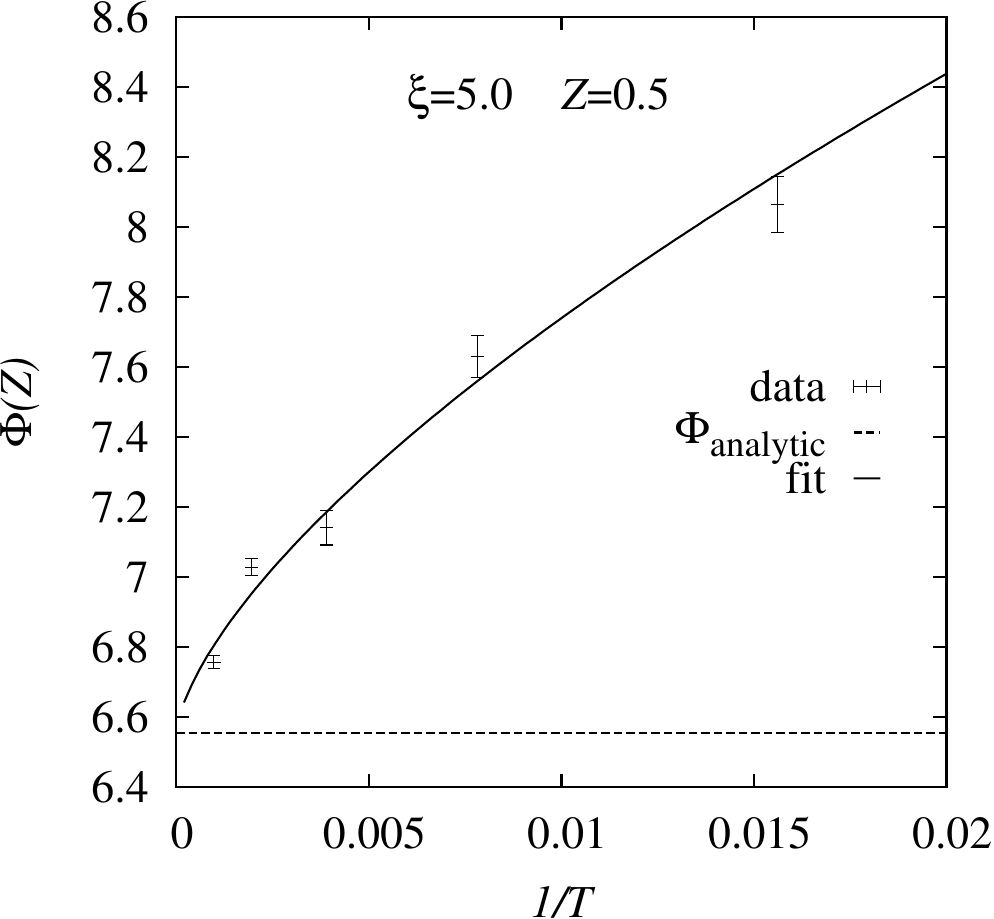}
    \caption{Extrapolation of the rate function for $\xi=5$ 
    to infinite walks lengths $T\to\infty$
    by showing the numerical result of $\Phi(Z)$ as function of $1/T$ for a fixed value of $Z=0.5$ and $T=\{ 64,128,256,512,1024\}$. The upper line
    shows the result of a fit according to Eq.~\eqref{eq:rate:fit}. The horizontal line indicates the analytic asymptotic value for $\Phi(Z=0.5)$. }
    \label{fig:extrapolating:rate}
\end{figure}

We have also performed a heuristic extrapolation by fitting the behavior as function of $T$, for a fixed value of $Z$, to a power law according to \begin{equation}
\Phi(Z,T)=\Phi_Z^{\infty} + a_Z T^{-b_Z}\,,
\label{eq:rate:fit}
\end{equation} 
i.e., with fitting parameters 
$\Phi^{\infty}$, $a$ and $b$ which may depend on $Z$. An example for such a fit is shown in Fig.~\ref{fig:extrapolating:rate}. As visible, the extrapolated value is compatible with the analytical result.

\subsection{Non convexity of $\Phi(Z)$ and first-order transition}

\begin{figure*}[!htb]
    \centering
    \includegraphics[width=0.46\linewidth]{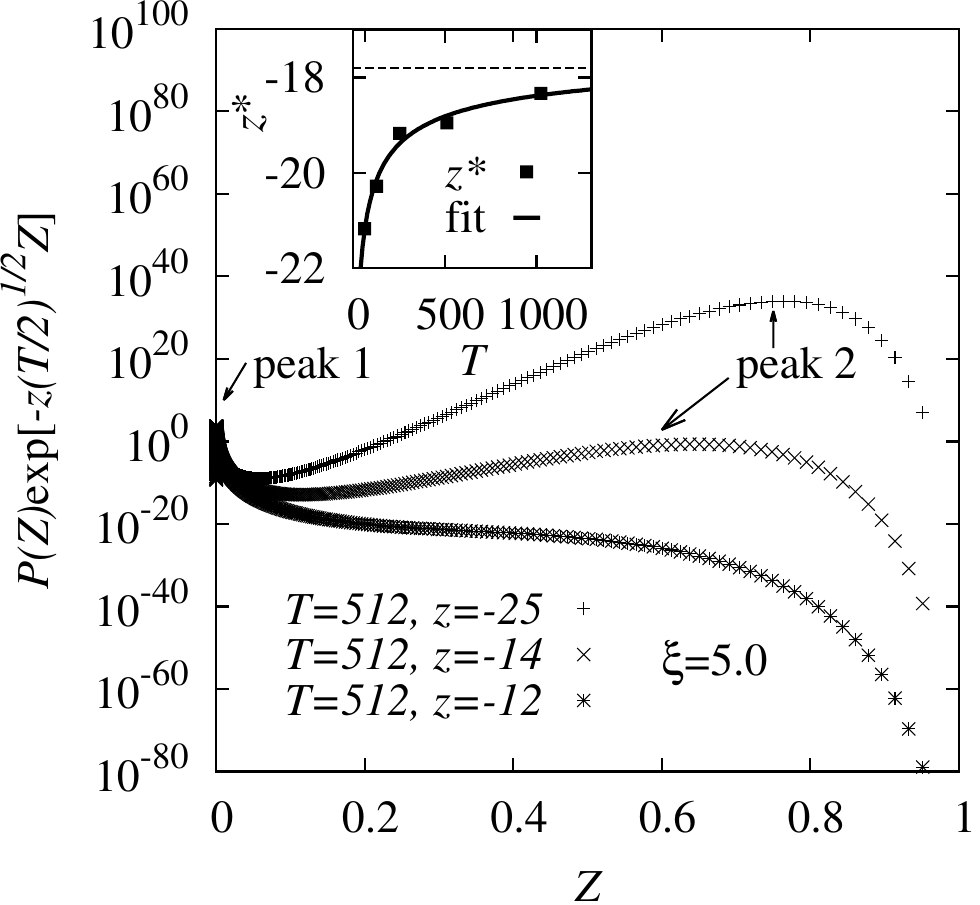}
    \hfill
    \includegraphics[width=0.45\linewidth]{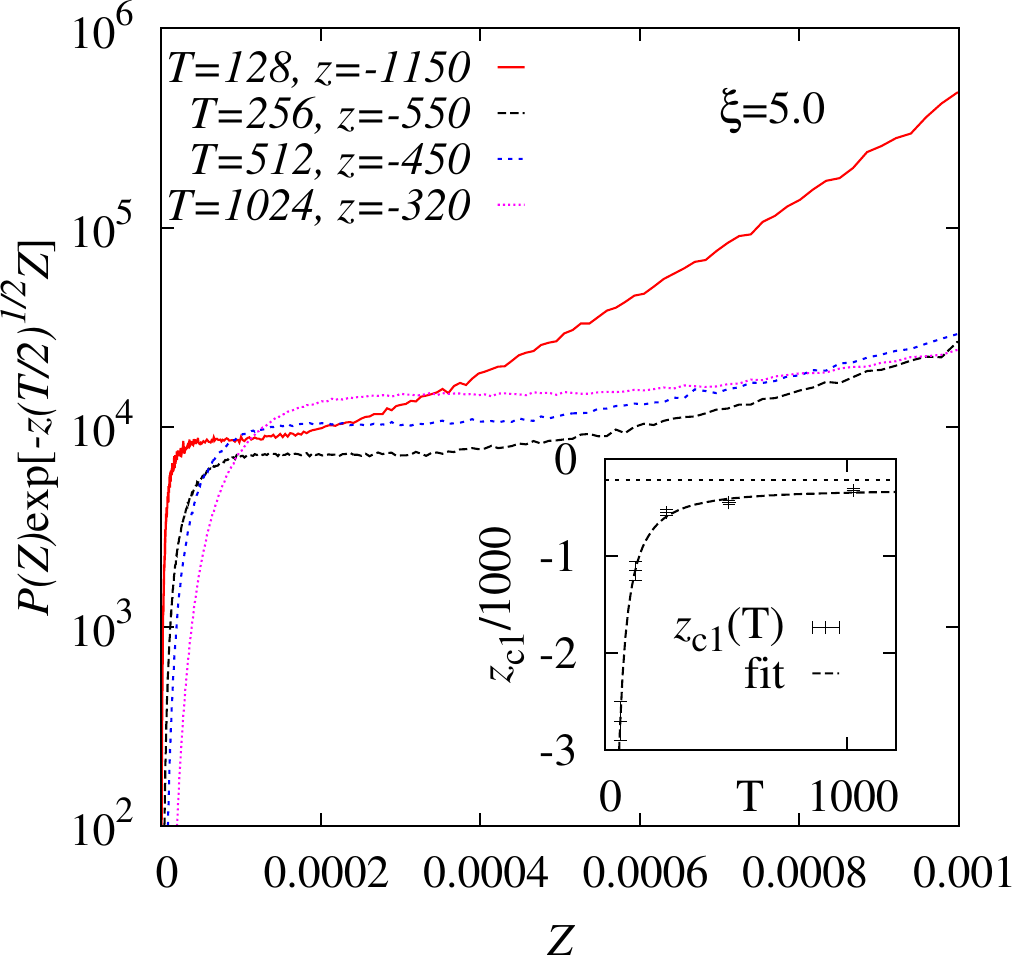}
    \caption{(left panel) Distributions $P(Z)$ rescaled with the factor $\exp(-z (T/2)^{1/2}Z)$. For intermediate values of $z$, the distribution 
    is observed to exhibit two peaks, as predicted. The inset shows the dependence of the scaling value $z^*$ where the two peaks attain the same height as function $T$ and a fit to a shifted power law, see text. The horizontal line shows the predicted value $z^*\simeq -17.84$. (right panel) For very negative values of $z\le z_{c1}(T)$, one observes  
    that the first peak close to $Z=0$ is suppressed. The inset illustrates the convergence to the predicted value $z_{c1}=-216.5$ at large $T$, which is included as horizontal line.
    }
    \label{fig:rescaled:rate}
\end{figure*}

A remarkable prediction of Ref.~\cite{KrajLedouCrossover} is that for the continuum model the rate function 
$\hat \Phi_\xi(Z)$ in \eqref{prediction1} is non-convex for $\xi >\sqrt{8}$. 
This results in a first-order phase transition in its Legendre transform (i.e. $\Psi_{\rm opt}(z)$ defined
in \eqref{Psiopt0}) 
associated to a tilted version of $P(Z)$ (see definition below and in Eq.~\eqref{tiltedm}). For a detailed
discussion see Appendix~\ref{subsec:multivaluation}, and for an illustration of the first-order
transition see Fig.~\ref{subfig:a}).
We thus predict that the same property holds for the Beta random walk, as we will now confirm.

We find that for large enough values of $\xi$, the numerical rate function $\Phi(Z)$ for the random walk exhibits a non-monotonic curvature, which is already visible in Fig.~\ref{fig:rate_fcts}. This leads to the appearance and disappearance of maxima in the tilted distribution 
$P(Z) \exp(-z(T/2)^{1/2}Z)$, depending on the choice of $z$. 
We recall that $\alpha=1$ here, hence this factor is also $\exp(-z \tilde T^{1/2}Z)$, 
corresponding to \eqref{tiltedm} for the continuum model.
Our prediction is thus that at large $T$ and for $\xi> \sqrt{8}$, $P(Z) \exp(-z(T/2)^{1/2}Z) $ should exhibit two peaks for $z \in [z_{c1},z_{c2}]$
given in \eqref{zc2}. For $\xi=5$ this corresponds to $z \in [-216.5,-13]$. The two peaks should become of same height 
for $z=z^*$ given in \eqref{zstar}, which for $\xi=5$ evaluates to $ z^* \simeq  - 17.84 $.

% when rescaling $\Phi(Z)$ with $\exp(-z(T/2)^{1/2}Z)$, depending on the choice of $z$. 
% \blue{We recall that $\alpha=1$ here, hence this factor is also $\exp(-z \tilde T^{1/2}Z)$
% }. The prediction is that $P(Z)$ should exhibit two peaks for $z \in [z_{c1},z_{c2}]$
% which for $\xi=5$ gives $z \in [-216.5,-13]$. The two peaks should become of same height 
% for $z=z^*= $

In the numerical results, we indeed observe that, 
for not too negative values of $z$, the rescaled rate functions exhibits a peak close to $Z=0$, see left of Fig.~\ref{fig:rescaled:rate}
the case $z=-12$. For intermediate values of $z$, 
a second peak appears, see the case $z=-14$. This is consistent with our analytical prediction recalled
above that a second peak should appear for $z \simeq -13$. This peak becomes slightly more pronounced when increasing the number $T$ of steps (not shown), and much more pronounced when decreasing the value of $z$.

For each value of $T$, there is a value $z^*(T)$ where both peaks exhibit the same height. The inset of 
left of Fig.~\ref{fig:rescaled:rate} shows $z^*(T)$
together with a fit to a power law $z^*(T)=z^*_{\infty}+a^* T^{-b^*}$, which results in $z^*_{\infty}=-17(1)$ which is well compatible
with the predicted value $z^*\simeq -17.84$.

For very negative values of $z$, the first peak disappears, see right of Fig.~\ref{fig:rescaled:rate}.
The value of $z$ above which this happens, which we predict to be $z_{c1}=-216.5$, is observed to be indeed very negative for small values of $T$, about  $z=-1150$ for $T=128$ and increases when increasing $T$, to
about $z=-320$ for $T=1024$.  When fitting $z_{c1}(T)$ to a heuristic power law of the form $z_{c1}(T)=z_{\textrm{c1}}^{\infty} + a_z T^{-b_z}$ we
obtain a limiting value $z_{\textrm{c1}}^{\infty}=-312(50)$ which is in rough agreement, i.e. within two sigma, with the limiting value $z_{c1}=-216.5$ (see inset). 

The behavior of
the numerically determined tilted PDF of $Z$ is thus in agreement with the prediction, and the first-order transition in
$\Psi_{\rm opt}(z)$ results when the second peak becomes higher than the first one (see Appendix~\ref{subsec:multivaluation}).

% \subsubsection{Point to point directed polymer mapping in a half-space with a hard-wall $A=+\infty$}
%   % \begin{figure}[ht!]
%   %   \centering
%   %   \includegraphics[scale=0.45]{lattice_polymer.eps}
%   %   \caption{Square lattice designed for the half-space problem with a hard-wall. The dashed lines are forbidden edges for the polymer constraining it to stay on the right of the lattice. An example of a polymer realization is drawn on the blue line. In this representation, we have $L=2\ell$.}
%   %   \label{fig:lattice_hardwall}
%   % \end{figure}
% To mimic a hard-wall type problem, one can forbid the polymer to visit some edges and introduce a tabu list for edges bringing the polymer towards the diagonal of the lattice. Indeed, the partition function should be strictly zero on the diagonal so that the probability of crossing the diagonal is strictly zero. We introduce this idea of forbidden edges in Fig.~\ref{fig:lattice_hardwall}. Finally, let us mention that an extension of this construction to a generic value of $A$ would be interesting.
% %To construct the directed polymer on the lattice and introduce a half-space like behavior, one could for instance modify the random weights of the lattice and still allow the polymer to visit the whole lattice. Indeed, if one thinks of the directed polymer in a continuum half-space problem with a symmetric wall ($A=0$), the mapping can be obtained by taking symmetric weights around the diagonal of the lattice
% %\begin{equation}
% %V_{x,t}=V_{-x,t}
% %\end{equation}
% %On the contrary, t

\section{Conclusion and outlook}

To summarize we have studied analytically and numerically the Beta random walk, a discrete time random walk on the square lattice
with Beta distributed time dependent i.i.d. jump probabilities with parameter $\alpha$. We have focused on the probability $Z$ that a walk
starting from the origin is at large time $T$ at position to the right of $X=\xi \sqrt{T/2}$ for a given $\xi>0$. 
We have determined analytically and numerically the law of large deviations of the observable $Z$. 
We have first predicted that the large-deviation rate function of the Beta random walk is
identical, up to a scale factor involving $\alpha$ that we determined, to the one of the continuum model for diffusion
in random media, which we recently obtained analytically (and is $\xi$-dependent).
This prediction was based on the large time asymptotic
analysis of an exact Fredholm determinant formula which exists for both the discrete and
the continuum problem. The prediction holds for any value of the parameter $\alpha$, hence it hints
at some universality in the large-deviation rate functions. Proving this prediction 
rigorously remains an open question for mathematicians.

In the absence of a rigorous proof, we performed a numerical test of this prediction.
We used a large-deviation sampling approach to measure the PDF $P(Z)$ for various
values of $\alpha$ and $\xi$. We were able to meaure the PDFs over many decades down to values as small as $10^{-100}$ and below. We obtained an accurate determination of the rate function
and observed convergence at large $T$ to the predicted analytical value. 
In addition, for a deeper investigation of the system properties beyond the overall shape of the distributions, we observed a first-order transition in the rate function $\Psi_{\rm opt}(z)$,
which manifests itself as multiple peaks in the tilted PDF of $Z$,
as predicted in Ref.~\cite{KrajLedouCrossover}. 
Our numerical results are thus also a confirmation of the predictions obtained
in that work.
The numerical methods used here should be useful to study 
the large-deviation regime for various models of 
diffusion of the extremal particle in a cloud of many random walkers
\cite{hass2023anomalous}.

\acknowledgments 
We thank G. Barraquand for helpful discussions. The simulations were performed at the HPC Cluster CARL, located at the University of Oldenburg (Germany) and funded by the DFG through its Major Research Instrumentation Program (INST 184/157-1 FUGG) and the Ministry of Science and Culture (MWK)
of the Lower Saxony State.
\newpage
\onecolumngrid
\appendix

\section{Analytical results}
\label{app:analytical}

We recall in this Appendix the analytical result of \cite{KrajLedouCrossover} for the continuum
model for arbitrary $\xi \geq 0$. To simplify notations in this Appendix all subscripts $\xi$ are implicit and $\tilde z$
is denoted $z$. We also recall that $Z=\tilde Z$ and $H=\tilde H$, hence we use
only the notations $Z$ and $H$ in place of $\tilde Z$ and $\tilde H$.\\

\begin{figure*}[h!]
    \centering
    \stackinset{l}{1.1cm}{b}{1.6cm}{\includegraphics[scale=0.2]{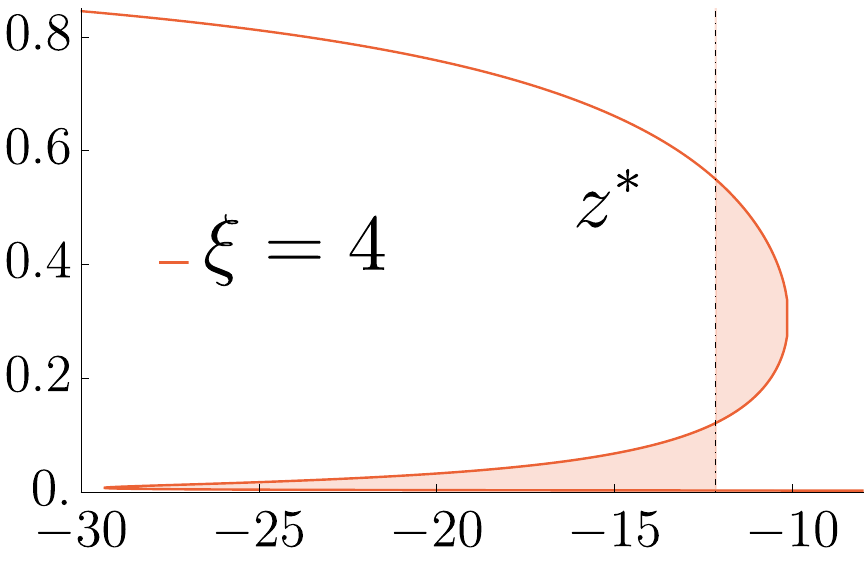}}{\includegraphics[scale=0.55]{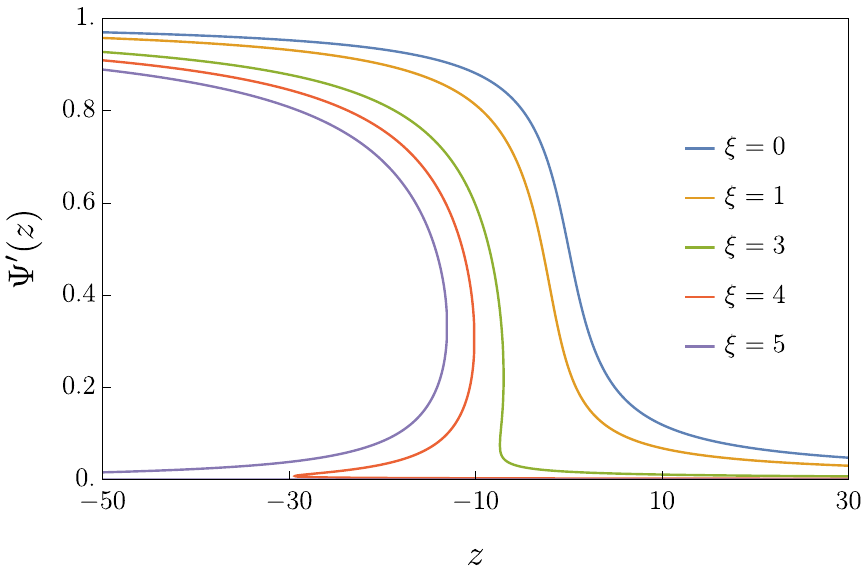}}
    \caption{For $\xi=(0,1,2,3,4,5)$ we plot the derivative rate function $\Psi'(z)$ from Table~\ref{tab:MainTextTableJump} as a function of $z$,
    with $\Psi'(+\infty)=0$ and $\Psi'(-\infty)=1$ (all the branches are shown).
    For $\xi > \xi_1$ and $z \in [z_{c1},z_{c2}]$ the function is multi-valued (see text).
    ({\bf Inset}) First-order transition: at $z=z^*$ such that the areas of the two shaded regions become equal
    the value of (the optimal) $\Psi_{\rm opt}'(z)$ (see definition in \eqref{Psiopt}) from one branch to the other, shown here for $\xi=4$.}
    \label{subfig:a}
\end{figure*}

The rate functions $\hat \Phi(Z)=\Phi(H)$ (with $H=\log Z$) 
is obtained quite generally from the parametric representation 
\be 
\begin{cases}
\hat \Phi(Z)= \Psi(z)- z Z ,\\
 Z = \Psi'(z)\, .
\end{cases}
\ee 
While the rate function $\hat \Phi(Z)$ is well defined and single valued,
for general $\xi>0$, $\Psi(z)$ may have several branches. This 
can be seen in Fig.~\ref{subfig:a} where in some cases one value
of $z$ corresponds to one or three values of $Z= \Psi'(z)$. \\

Below, we first explain how to compute $\Psi(z)$ and its various
branches, and then we explain how to perform the Legendre inversion.
Finally we discuss the multivaluation and the 
first-order transition
of the "optimal" $\Psi_{\rm opt}(z)$, see Fig.~\ref{subfig:a}.

\subsection{How to compute $\Psi(z)$}

One first defines 
\be \label{PsiFinal0} 
\begin{split}
\Psi^0(z) &= -
\dashint_\R \frac{\rmd q}{2 \pi}\frac{\mathrm{Li}_2( z (\I q - \frac{\xi}{2}) e^{-q^2 - \frac{\xi^2}{4}} )}{(\I q - \frac{\xi}{2})^2} % +z\Theta(-\xi)\\
\\
\end{split}
\ee 
The general formula for
$\Psi(z)$ takes the form
\be \label{continuation1} 
\Psi(z) = \Psi^0(z) + \Delta(z) 
\ee 
where $\Psi^0(z)$ is the same integral as in \eqref{PsiFinal0}. 
Note that we compute $\Psi^0(z)$ numerically using the default \texttt{PolyLog} function in Mathematica together with the \texttt{NIntegrate} routine for integration (with in some cases a branch cut on the integration contour dealt automatically by Mathematica).
The convention $\Delta(z)=0$ defines the main branch of $\Psi(z)$. The other branches and the form of $\Delta(z)$ as a function of $\xi$ and $z$ are
shown in Table~\ref{tab:MainTextTableJump}. 
Note that for $z \geq 0$ one has $\Delta(z)=0$.\\ 

\begin{table*}[t!]
    \centering
    \begin{tabular}{c|c|c|c }
            \hline 
        \hline 
        &&&\\[-0.8ex]
         $\xi$ & $ 0\leq \xi \leq \xi_1$ & \parbox{2cm}{$\xi_1\leq \xi \leq \xi_2$\\ $z_{c1}<z_{c2}<z_c$} & \parbox{2cm}{$ \xi_2\leq \xi$\\$z_{c1}<z_c<z_{c2}$}\\[2ex]
        \hline &&&\\[-0.5ex]
        $\Delta(z)=$  & 
        $\begin{cases}
        0, &z_c<z \\
        \Delta_1(z),  &z<z_c 
        \end{cases}
        $
        &
       $\begin{cases}
        0, &z_c<z \\
        \Delta_1(z), &z_{c1}<z<z_c\\
        \Delta_2(z), \,  &z_{c1}<z<z_{c2}\\
        \Delta_3(z), \, & z<z_{c2}
        \end{cases}$ 
        & 
        $\begin{cases}
        0, &z_c<z \\
        \Delta_1(z), &z_{c1}<z<z_c\\
        \Delta_2(z), \,  &z_{c1}<z<z_{c}\\
        \Delta_2(z)-\Delta_1(z), \,  &z_{c}<z<z_{c2}\\
        \Delta_3(z)-\Delta_1(z), \,  &z_{c}<z<z_{c2}\\
        \Delta_3(z), \, & z<z_{c}
        \end{cases}$ \\[1ex]
            \hline 
        \hline
    \end{tabular}
    \caption{Determination of the jump function $\Delta(z)$ in the different phases in the case $\xi\geq 0$. 
    One has $z_c= - \frac{2}{\xi} e^{\xi^2/4} \leq 0$ and the points $z=z_{c1}$ and $z=z_{c2}$ are turning points
    which depend on $\xi$. In the interval $z\in [z_{c1},z_{c2}]$, the function $\Delta(z)$ is multi-valued (i.e. it has several branches) 
    due to these turning points. The definition of $\Delta_\ell$ is given in \eqref{De1}.}
    \label{tab:MainTextTableJump}
\end{table*}

The jump functions $\Delta_\ell(z)$ for $\ell=\{1,2,3 \}$ which appear in this Table are defined as follows.
First one has
 \bea \label{De1} 
 \Delta_\ell(z) = \hat \Delta(p_\ell(z,\xi))
\eea 
where
\begin{equation}
\label{defhatDe}
\begin{split}
    \hat \Delta(p) =& 
\frac{1}{\xi}\bigg[-(\xi ^2+2) (\log (\xi )-\log (\xi +2 p))+2 p (p-\xi )-\frac{4 p}{\xi +2 p}\bigg] 
\end{split}
\end{equation}

The $p_\ell(z,\xi)$ are the real roots of the equation for $p$
\begin{equation}
\label{eq:BranchCutEquation}
 e^{- p^2 + \frac{\xi^2}{4}} + z (p + \frac{\xi}{2})=0   \, .
\end{equation}

The behavior of these roots is as follows.

Let us define $z_c =-\frac{2}{\xi} e^{\frac{\xi ^2}{4}}$. 
For $z_c\leq z\leq 0$, there is always one positive zero to \eqref{eq:BranchCutEquation} denoted $p_1=p_{1}(z,\xi)$. For $z < z_c$, the zeroes of \eqref{eq:BranchCutEquation} are all negative and their number is:
\begin{enumerate}
    \item for $0<\xi<\xi_1=\sqrt{8}$, there is one zero $p_{1}(z,\xi)$;
    \item for $\xi_1<\xi$ and $z \in ]z_{c1},z_{c2}[$ there are three zeroes $p_{1}(z,\xi)>p_{2}(z,\xi)>p_{3}(z,\xi)$. 
    The zeroes degenerate, i.e. $p_{1}=p_{2}$ for $z=z_{c1}$ and $p_{2}=p_{3}$ for $z=z_{c2}$ which define $z_{c1},z_{c2}$.
    For $z>z_{c2}$, there is only one zero $p_{1}(z,\xi)$. For $z < z_{c1}$, there is only one zero $p_{3}(z,\xi)$. 
\end{enumerate}
Note that $z_{c1}<z_{c2}<0$, with $z_{c1}=z_{c2}$ at $\xi=\xi_1$, 
and their explicit expression and dependence on $\xi$ is given 
for $\xi> \xi_1=\sqrt{8}$ by
\be
\begin{split}
&    z_{c1}=-\frac{1}{2} e^{\frac{1}{8} \left(\xi  \left(\xi+\sqrt{\xi ^2-8} \right)+4\right)} \left(\xi -\sqrt{\xi ^2-8}\right)
 \\
&   z_{c2}=-\frac{1}{2} e^{\frac{1}{8} \left(\xi \left(\xi-\sqrt{\xi ^2-8}\right)  +4\right)} \left(\xi+\sqrt{\xi ^2-8} \right)
 \label{zc2} 
 \end{split}
\ee
Note that $z_c$ and $z_{c2}$ become equal at the value $\xi=\xi_2$ with
\be 
\begin{split}
   \xi_2 &=-2 \sqrt{\frac{2}{-2 W_{-1}\left(-\frac{1}{2 \sqrt{e}}\right)-1}} W_{-1}\left(-\frac{1}{2 \sqrt{e}}\right)\\
   &\simeq 3.13395\\
   \end{split}
\ee
where $W_{-1}$ is the Lambert function \cite{corless1996lambertw}.

\subsection{Inversion of Legendre transform }
\label{sec:supp-mat-table-legendre}

Defining the critical height $H_c=\log \Psi_0'(z_c)$, the rate function $\Phi(H)$ is given 
by the parametric representation displayed in Table~\ref{tab:app-case2} for $\xi \leq \xi_1=\sqrt{8}$.
\begin{table*}[h!]
    \centering
    \begin{tabular}{p{2cm} p{2cm} p{3.5cm}  p{5cm} }
        \hline 
        \hline 
        &&&\\[-0.8ex]
        interval of $H$ & interval of $z$ & $H=$ & $\Phi(H)=$ \\[0.5ex]
        \hline \\[-0.5ex]
        $H\leq H_c$ & $z_c\leq z$ &$\log \Psi_0'(z)$ & $\Psi_0(z)- z \Psi_0'(z)$\\[2ex]
        $0 > H> H_c$  & $z_c> z$ &$ \log (\Psi_0'(z)+\Delta_1'(z)) $ & $\Psi_0(z)+\Delta_1(z)- z (\Psi_0'(z)+\Delta_1'(z))$\\[1ex]
            \hline 
        \hline
    \end{tabular}
    \caption{Case $0<\xi \leq \xi_1$}
    \label{tab:app-case2}
\end{table*}

For $\xi_1 < \xi \leq \xi_2$ it is given by the parametric representation displayed in Table~\ref{tab:app-case3}
where we have defined
\begin{equation}
\begin{split}
    H_{c1}&=\log (\Psi_0'(z_{c1})+\Delta_1'(z_{c1}))=\log (\Psi_0'(z_{c1})+\Delta_2'(z_{c1})),\\
    H_{c2}&=\log (\Psi_0'(z_{c2})+\Delta_2'(z_{c2}))=\log (\Psi_0'(z_{c2})+\Delta_3'(z_{c2})),
\end{split}
\end{equation}

\begin{table*}[h!]
    \centering
    \begin{tabular}{p{3cm} p{2.5cm} p{3.5cm}  p{5cm} }
    \hline 
        \hline 
        &&&\\[-0.8ex]
        interval of $H$ & interval of $z$ & $H=$ & $\Phi(H)=$ \\[2ex]
        \hline &&&\\[-0.5ex]
        $H\leq H_c$ & $z_c\leq z$ &$\log \Psi_0'(z)$ & $\Psi_0(z)- z \Psi_0'(z)$\\[2ex]
        $H_c<H\leq H_{c1}$  & $z_{c1}\leq z<z_c$ &$ \log (\Psi_0'(z)+\Delta_1'(z)) $ & $\Psi_0(z)+\Delta_1(z)- z (\Psi_0'(z)+\Delta_1'(z))$\\[2ex]
        $H_{c1}<H\leq H_{c2}$  & $z_{c1}< z\leq z_{c2}$ &$ \log (\Psi_0'(z)+\Delta_2'(z)) $ & $\Psi_0(z)+\Delta_2(z)- z (\Psi_0'(z)+\Delta_2'(z))$\\[2ex]
        $H_{c2}<H<0$  & $z_{c2}> z$ &$ \log (\Psi_0'(z)+\Delta_3'(z)) $ & $\Psi_0(z)+\Delta_3(z)- z (\Psi_0'(z)+\Delta_3'(z))$\\[1ex]
            \hline 
        \hline
    \end{tabular}
    \caption{Case $\xi_1< \xi \leq \xi_2$}
    \label{tab:app-case3}
\end{table*}

For $\xi_2 < \xi $ it is given by the parametric representation displayed in Table~\ref{app-tab-branches-large-xi}
where we have defined
\begin{equation}
\begin{split}
    H_{c10}&=\log (\Psi_0'(z_{c1})+\Delta_1'(z_{c1})),\\
    H_{c11}&=\log (\Psi_0'(z_{c})+\Delta_2'(z_{c})),\\
    H_{c20}&=\log (\Psi_0'(z_{c2})+\Delta_2'(z_{c2})-\Delta_1'(z_{c2})),\\
    H_{c21}&=\log (\Psi_0'(z_{c})+\Delta_3'(z_{c})),
\end{split}
\end{equation}

\begin{table*}[h!]
    \centering
    \begin{tabular}{p{3cm} p{2.3cm} p{3.8cm}  p{7cm} }
    \hline 
        \hline 
        &&&\\[-0.8ex]
        interval of $H$ & interval of $z$ & $H=$ & $\Phi(H)=$ \\[2ex]
        \hline &&&\\[-0.5ex]
        $H\leq H_c$ & $z_c\leq z$ &$\log \Psi_0'(z)$ & $\Psi_0(z)- z \Psi_0'(z)$\\[2ex]
        $H_c<H\leq H_{c10}$  & $z_{c1}\leq  z< z_{c}$ &$ \log (\Psi_0'(z)+\Delta_1'(z)) $ & $\Psi_0(z)+\Delta_1(z)- z (\Psi_0'(z)+\Delta_1'(z))$\\[2.5ex]
        $H_{c10}<H\leq H_{c11}$  & $z_{c1}< z\leq z_{c}$ &$ \log (\Psi_0'(z)+\Delta_2'(z)) $ & $\Psi_0(z)+\Delta_2(z)- z (\Psi_0'(z)+\Delta_2'(z))$\\[2.5ex]
        $H_{c11}<H\leq H_{c20}$  & $z_{c}< z\leq z_{c2}$ &$  \log (\Psi_0'(z)+\Delta'_2(z)-\Delta_1'(z)) $ & $\Psi_0(z)+\Delta_2(z)-\Delta_1(z)- z (\Psi_0'(z)+\Delta_2'(z)-\Delta_1'(z))$\\[2.5ex]
        $H_{c20}<H\leq H_{c21}$  & $z_{c}\leq  z< z_{c2}$ &$ \log (\Psi_0'(z)+\Delta'_3(z)-\Delta_1'(z))  $ & $\Psi_0(z)+\Delta_3(z)-\Delta_1(z)- z (\Psi_0'(z)+\Delta_3'(z)-\Delta_1'(z))$\\[2.5ex]
        $H_{c21}<H<0$  & $z_{c}> z$ &$ \log (\Psi_0'(z)+\Delta_3'(z)) $ & $\Psi_0(z)+\Delta_3(z)- z (\Psi_0'(z)+\Delta_3'(z))$\\[1ex]
            \hline 
        \hline
    \end{tabular}
    \caption{Case $ \xi_2< \xi$}
    \label{app-tab-branches-large-xi}
\end{table*}

\subsection{Multi-valuation and first-order transition}
\label{subsec:multivaluation} 

To interpret the $S$-shape form of $\Psi'(z)$ shown with all its branches
in Figure~\ref{subfig:a}, we recall the definition of the "optimal"
$\Psi(z)$ defined as
\be \label{Psiopt0}
\Psi_{\rm opt}(z) = \min_{Z \in [0,1]}[\hat \Phi(Z) + z Z ] 
\ee 
It has the property that its derivative obeys
\be 
\Psi_{\rm opt}'(z) = \langle Z \rangle_z
\ee 
where
$\langle Z \rangle_z$ is the expectation value for large $\tilde T$ of the random
variable $Z$ under the $z$-dependent tilted measure
\be \label{tiltedm} 
P(Z) e^{- \sqrt{\tilde T}  z Z} \sim e^{- \sqrt{\tilde T} (\hat \Phi(Z) + z Z) } 
\ee 
The key point is that for $\xi>\xi_1$ the function $\hat \Phi(Z)$ has a concave part \cite{KrajLedouCrossover}. %see Fig.~\ref{subfig:d}.
As a consequence, for $z \in [z_{c1},z_{c2}]$ the tilted measure \eqref{tiltedm} develops three extrema at $Z_j(z)=e^{H_j(z)}$,
solutions of $\hat \Phi'(Z)=-z$. They lead to the
three branches of $\Psi'(z)=Z_j(z)$. 
% Equivalently, there are 
% three extremal values $H_j(z)$ in \eqref{Legendre0} solutions of \eqref{eq:ParametricRepresentation}.
The "optimal" $\Psi_{\rm opt}(z)$ is determined by the absolute minimum in
\eqref{Psiopt0} (which corresponds to the absolute maximum in the tilted PDF of $Z$) 
hence it is given by
\be \label{Psiopt}
\Psi_{\rm opt}(z) =  \min_{j=1,2,3} [\hat \Phi(Z_j) + z Z_j ]
\ee 
and the optimal $j$ switches from $j=1$ to $j=3$ at $z=z^*(\xi)$ where $z^*$ is the solution of 
\be \label{zstar} 
\Delta_1(z^*)=\Delta_3(z^*) \, .
\ee 
It is also the point given by an equal area law on the curve $\Psi'(z)$, as in standard
magnetization versus field curve for a first-order phase transition, 
see Fig.~\ref{subfig:a} (inset).
The points $Z=\{Z_1, Z_3 \}$ are "stable" whereas $Z=Z_2$ is "unstable". The optimal rate function $\Psi_{\rm opt}(z)$ thus exhibits a first-order transition. This type
of transition occurs in other large-deviation problems
\cite{TouchetteReview2018}.

\section{Technical details of the importance sampling algorithm} 
\label{app:sec:technical-details}

 To sample a wide range of values of $H$, one chooses a
suitable set of parameters $\{\theta_{-N_{n}},\theta_{-N_{n}+1},\ldots,
\theta_{N_{p}-1},\theta_{N_{p}}\}$,
$N_{n}$ and $N_{p}$ being the number of negative
and positive parameters, to access the large-deviation regimes (left and right). 
The normalisation constants $W(\theta_i)$ are obtained 
by first computing the histogram using direct sampling, corresponds to $\theta=0$.
Then for $\theta_{+1}$, one matches the right part of the biased histogram with the left tail of the unbiased one and for $\theta_{-1}$, one matches the left part of the biased histogram with the right tail of the unbiased one. Similarly one iterates for the other values of $\theta$ and the corresponding \emph{relative} normalisation
constants  can be obtained. In the end the full distribution is normalized to result in a total probability of one.\\

%\end{widetext} 

%%%%% CLEAR DOUBLE PAGE!
\newpage{\pagestyle{empty}\cleardoublepage}

\end{document}